\RequirePackage{fix-cm}
\documentclass{svjour3}                     
\AtBeginDocument{ \paperwidth=\dimexpr 1in + \oddsidemargin + \textwidth + 1in + \oddsidemargin \relax \paperheight=\dimexpr 1in + \topmargin + \headheight + \headsep + \textheight + 1in + \topmargin \relax \usepackage[pass]{geometry} \relax }
\usepackage{graphicx, color}
\usepackage{hyperref}
\usepackage{natbib}
\usepackage[figuresleft]{rotating}
\usepackage{caption}
\usepackage{subcaption}
\usepackage{layout}
\usepackage{setspace}
\usepackage{lscape}
\usepackage{dcolumn}
\usepackage{booktabs}
\usepackage[T1]{fontenc} 
\usepackage{tgtermes, tgheros} 
\usepackage{amsmath}
\usepackage{amssymb} 
\usepackage{autonum}
\usepackage{mathtools} 
\usepackage{bm}
\usepackage{multirow}
\usepackage{threeparttablex} 
\usepackage{array}
\usepackage{tabularx}
\usepackage{longtable}
\usepackage{commath}
\usepackage{comment}
\usepackage[table]{xcolor}
 \usepackage{mathrsfs} 

\journalname{ }
\def\makeheadbox{\relax}

\begin{document}
\title{ 
The elastic origins of tail asymmetry
}
\titlerunning{}        
\author{
Satoshi Nakano \and Kazuhiko Nishimura
}
\institute{
Satoshi Nakano \at Nihon Fukushi University, 477-0031 Japan\\ 
\email{nakano@n-fukushi.ac.jp} 
\and Kazuhiko Nishimura \at Chukyo University, 466-8666 Japan \\
\email{nishimura@lets.chukyo-u.ac.jp} 
}
\date{\today}
\maketitle
\begin{abstract}
Based on a multisector general equilibrium framework, we show that the sectoral elasticity of substitution plays the key role in the evolution of asymmetric tails of macroeconomic fluctuations and the establishment of robustness against productivity shocks.
A non-unitary elasticity of substitution renders a nonlinear Domar aggregation, where normal sectoral productivity shocks translate into non-normal aggregated shocks with variable expected output growth. 
We empirically estimate 100 sectoral elasticities of substitution,
using the time-series linked input-output tables for Japan, and find
that the production economy is elastic overall, relative to a
Cobb-Douglas economy with unitary elasticity.  
In addition to the previous assessment of an inelastic production economy for the US, the contrasting tail asymmetry of the distribution of aggregated shocks between the US and Japan is explained.
Moreover, the robustness of an economy is assessed by expected output growth, the level of which is led by the sectoral elasticities of substitution under zero-mean productivity shocks.
\end{abstract}

\clearpage
\section{Introduction}
The subject of how microeconomic productivity shocks translate into aggregate macroeconomic fluctuations, in light of production networks, has been widely studied in the business cycle literature.
Regarding production networks, the works of \citet{lpJPE}, \citet{horvath98, horvath}, and \citet{dupor} are concerned with input-output linkages, whereas \citet{aceECTA, aceAER, ace2020} base their analysis on a multisectoral general equilibrium model under a unitary elasticity of substitution, or Cobb-Douglas economy. 
In a Cobb-Douglas economy, Domar aggregation becomes linear with respect to sectoral productivity shocks, and because the Leontief inverse that plays the essential role in their aggregation is granular \citep{gabaixECTA}, some important dilation of volatility in aggregate fluctuations becomes explainable.

\begin{figure}[t!]
\centering
\includegraphics[width=0.495\textwidth]{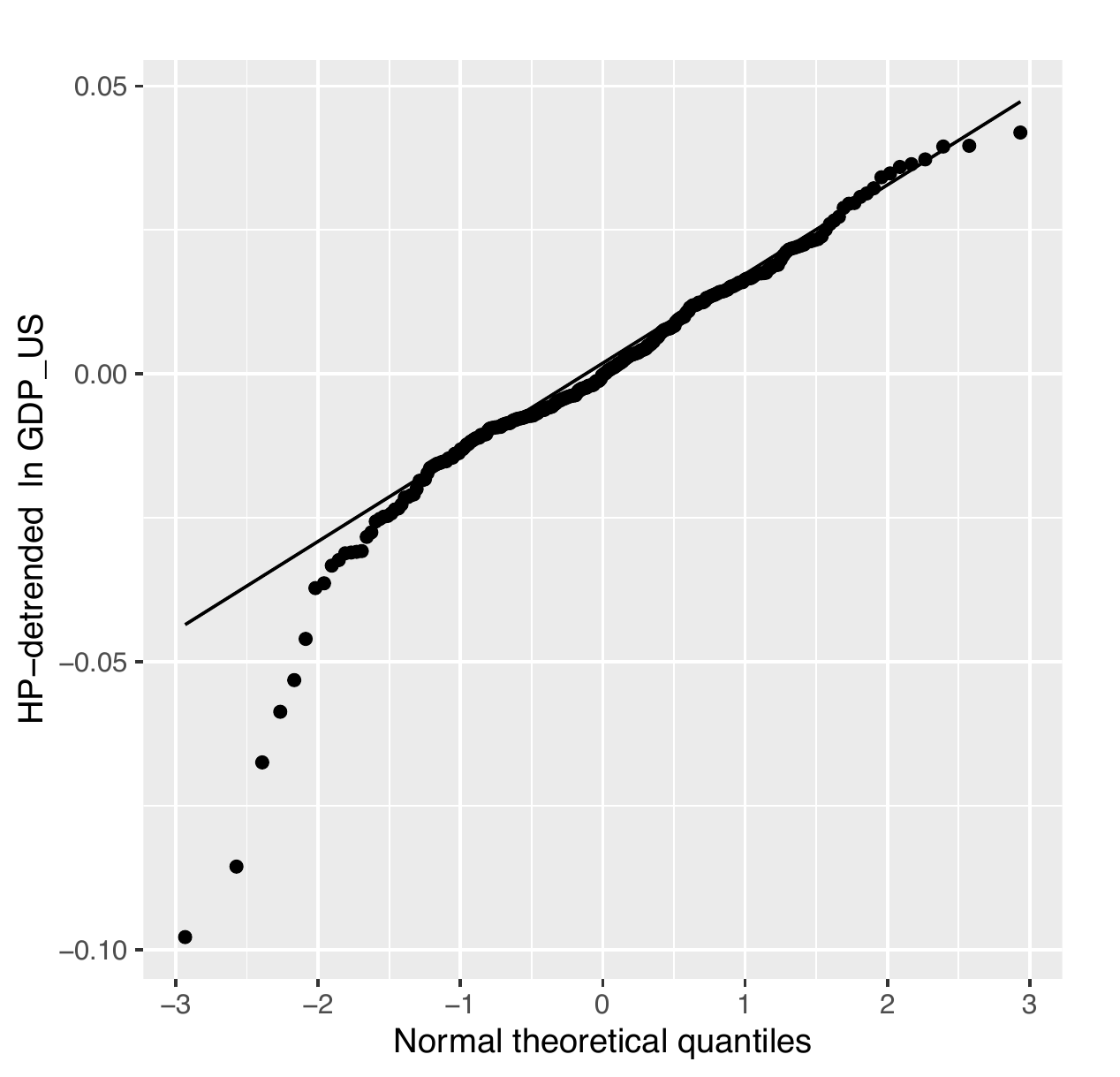}
\includegraphics[width=0.495\textwidth]{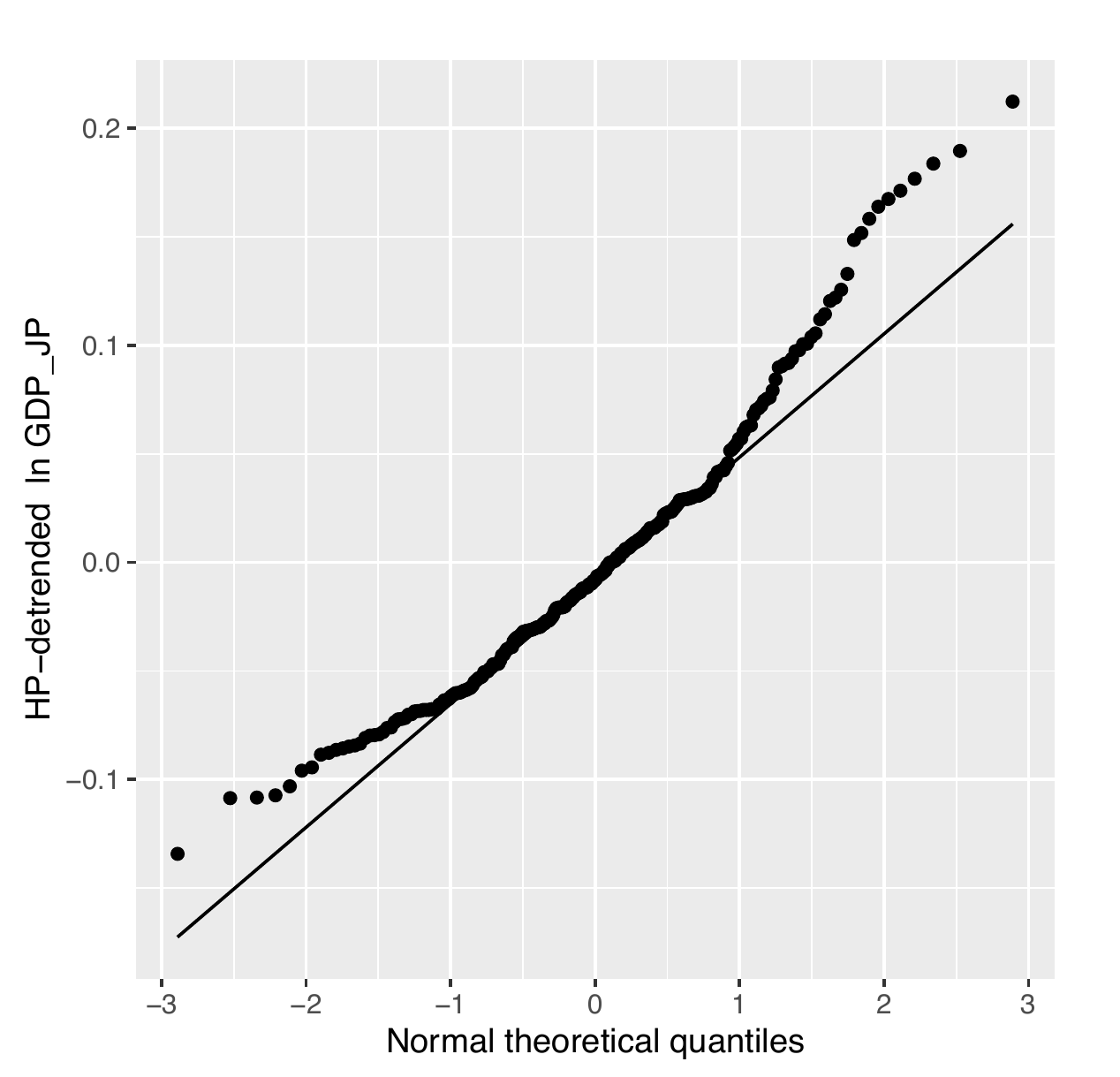}
\caption{
Quantile-quantile plots of postwar (US:1947I--2021II; Japan:1955II--2020I) quarterly HP-detrended log GDP against the normal distribution for the US (left) and Japan (right).
Source: Quarterly GDP data are taken from \citet{fred} and the \citet{co}.
} \label{fig0}
\end{figure}

Moreover, peculiar aggregate fluctuations are evident from the statistical record.
Figure \ref{fig0} depicts the quantile-quantile (QQ) plots of the
HP-detrended postwar quarterly log GDP using the Hodrick-Prescott (HP)
filter against the standard normal for the US (left) and Japan (right).
These figures indicate that either the Cobb-Douglas or the normal
shock assumption is questionable, since these assumptions together
make the QQ plot a straight line.
\citet{aceAER} explained the non-normal frequency of large economic downturns (in the US), using non-normal (heavy tailed) microeconomic productivity shocks.
\citet{baqaee}, on the other hand, claim that a non-Cobb-Douglas economy (thus, with nonlinear Domar aggregation) can lead to such non-normality in macroeconomic fluctuations under normally distributed productivity shocks.

Indeed, the asymmetric tails in the left panel of Figure \ref{fig0}
seem to coincide with the case in which the aggregated shocks are evaluated in a Leontief economy. 
We know this because a Cobb-Douglas economy with more alternative
technologies can always yield a better solution (technology) than a Leontief economy with a single technology.
Thus, if a Cobb-Douglas economy generates an aggregate output that corresponds to the straight line of the QQ plot, an unrobust Leontief economy that can generate less than a Cobb-Douglas economy must take the QQ plot below the straight line.
This feature (of an inelastic economy) 
is also consistent with the analysis based on the general equilibrium model with intermediate production with a very low (almost Leontief) elasticity of substitution studied in \citet{baqaee}.

If the theory that the elasticity of substitution dictates the shape of the tails of the distribution of the aggregate macroeconomic shocks were to stand, Japan would have to have an elastic economy according to the right panel of Figure \ref{fig0}.
Consequently, one basis of this study is to empirically evaluate the sectoral elasticity of substitution for the Japanese economy.\footnote{The role of elasticities in propagating shocks for multisector models with input-output linkages is also highlighted in \citet{carvalho_QJE} with regard to Japan's supply chain.}
To do so, we utilize the time-series linked input-output tables, spanning 100 sectors for 22 years (1994--2015), available from the \citet{jip} database.
We extract factor prices (as deflators) from the linked transaction tables available in both nominal and real values.
We use the sectoral series of TFP that are also included in the database to instrument for the potentially endogenous explanatory variable (price) in our panel regression analysis.
Note in advance that our sectoral average elasticity estimates ($\bar{\hat{\sigma}}=1.54$) exceeded unity.

To ensure that our study is compatible with the production networks across sectors, we construct a multisector general equilibrium model with the estimated sector-specific CES elasticities.
We assume constant returns to scale for all production so that we can work on the system of quantity-free unit cost functions to study the potential transformation of the production networks along with the propagation of productivity shocks in terms of price.
Specifically, given the sectoral productivity shocks, the fixed-point solution of the system of unit cost functions allows us to identify the equilibrium production network (i.e., input-output linkages) by the gradient of the mapping.
By eliminating all other complications that can potentially affect the
linearity of the Domar aggregation, we are able to single out the role
of the substitution elasticity on the asymmetric tails of the aggregated shocks. 

The remainder of this paper proceeds as follows.
We present our benchmark model of a CES economy with sector-specific elasticities and then reduce the model to Leontief and Cobb-Douglas economies in Section 2. 
We also refer to the viability of the equilibrium structures with respect to the aforementioned economies and show that non-Cobb-Douglas economies are not necessarily prevented from exhibiting an unviable structure.
In Section 3, we present our panel regression equation and estimate sectoral elasticities of substitution with respect to the consistency of the estimator.
Our main results are presented in Section 4 where we show that our nonlinear (and recursive) Domar aggregators for non-Cobb-Douglas economies qualitatively replicate the asymmetric tails presented in this section.
Section 5 concludes the paper.

\section{The CES Economy}
\subsection{Model}
Below are a constant-returns-to-scale CES production function and the
corresponding CES unit cost function for the $j$th sector (index
omitted) of $n$ sectors, with $i = 1, \cdots, n$ being an intermediate and a single primary factor of production labelled $i=0$.
\begin{align}
x = z \left( \sum_{i=0}^n (\alpha_{i})^{\frac{1}{\sigma}} (x_{i})^{\frac{\sigma-1}{\sigma}}  \right)^{\frac{\sigma}{\sigma-1}},
&&
\pi =\frac{1}{z}\left( \sum_{i=0}^n \alpha_{i} (\pi_{i})^{\gamma}  \right)^{{1}/{\gamma}} 
\end{align}
Here, $\sigma =1-\gamma$ denotes the elasticity of substitution, while $\alpha_{i}$ denotes the share parameter with $\sum_{i=0}^n \alpha_{i}=1$. 
Quantities and prices are denoted by $x$ and $\pi$, respectively.
Note that the output price equals the unit cost due to the constancy of returns to scale. 
The Hicks-neutral productivity level of the sector is denoted by $z$.
The duality asserts zero profit in all sectors $j=1,\cdots,n$, i.e., $\pi_j x_j = \sum_{i=0}^n \pi_i x_{ij}$. 

By applying Shephard's lemma to the unit cost function of the $j$th sector, we have:
\begin{align}
s_{ij} 
= \alpha_{ij} \left( \frac{z_j \pi_j}{\pi_i} \right)^{-\gamma_j}
\label{sij}
\end{align}
where $s_{ij}$ denotes the $i$th factor cost share of the $j$th sector.
For later convenience, let us calibrate the share parameter at the benchmark where price and productivity are standardized, i.e., $\pi_0 = \pi_1 = \cdots = \pi_n = 1$ and $z_1 = \cdots = z_n =1$. 
Since we know the benchmark cost share structure from the input-output coefficients of the benchmark period $a_{ij}$, the benchmark-calibrated share parameter must therefore be $\alpha_{ij} = a_{ij}$.
Taking this into account, the equilibrium price $\bm{\pi}=(\pi_1, \cdots, \pi_n)$ given $\bm{z}=(z_1, \cdots, z_n)$ must be the solution to the following system of $n$ equations:
\begin{align}
\pi_1 &= (z_1)^{-1} \left( 
a_{01} (\pi_0)^{\gamma_1} + a_{11} (\pi_1)^{\gamma_1} + \cdots + a_{n1} (\pi_n)^{\gamma_1} 
\right)^{1/\gamma_1} \\
\pi_2 &= (z_2)^{-1} \left( 
a_{02} (\pi_0)^{\gamma_2} + a_{12} (\pi_1)^{\gamma_2} + \cdots + a_{n2} (\pi_n)^{\gamma_2} 
\right)^{1/\gamma_2} \\
 &\vdots \\
\pi_n &= (z_n)^{-1} \left( 
a_{0n} (\pi_0)^{\gamma_n} + a_{1n} (\pi_1)^{\gamma_n} + \cdots + a_{nn} (\pi_n)^{\gamma_n} 
\right)^{1/\gamma_n}
\end{align}
where we can set the price of the primary factor $\pi_0$ as the num{\'e}raire.
For later convenience, we write this system in a more concise form as follows:
\begin{align}
\bm{\pi}=\left< \bm{z} \right>^{-1} \bm{c}\left(\bm{\pi}; \pi_0 \right)
\label{eq}
\end{align}
Here, angled brackets indicate the diagonalization of a vector.
Note that $\bm{c}: \mathbb{R}^{n}_{+} \to \mathbb{R}^{n}_{+}$ is strictly concave and $\bm{z} \in \mathbb{R}^{n}_{+}$.
Consider a mapping $\mathcal{E}$ that nests the equilibrium solution (fixed point) $\bm{\pi}$ of (\ref{eq}) and maps the (exogenous) productivity $\bm{z}$ onto the equilibrium price $\bm{\pi}$, i.e.,
\begin{align}
\bm{\pi}=\mathcal{E} (\bm{z}; \pi_0 )
\label{eqe}
\end{align}

There is no closed-form solution to (\ref{eqe}).
However, one can be found for the case of uniform elasticity, i.e., $\gamma_1 = \cdots = \gamma_n = \gamma$, which is as follows:
\begin{align}
\bm{\pi} = \pi_0 \left( \bm{a}_0 \left[ \left< \bm{z} \right>^\gamma  - \mathbf{A} \right]^{-1} \right)^{1/\gamma}
&&
\text{Uniform CES economy}
\label{cesuni}
\end{align}
where the $n$ row vector $\bm{a}_0 = (a_{01}, \cdots, a_{0n})$ is called the primary factor coefficient vector and the $n \times n$ matrix $\mathbf{A} = \left\{ a_{ij} \right\}$ is called the input-output coefficient matrix.
The case of a Leontief economy, where $1-\gamma =0$, (\ref{cesuni}), can be reduced straightforwardly as follows:
\begin{align}
\bm{\pi} = \pi_0  \bm{a}_0 \left[ \left< \bm{z} \right> - \mathbf{A} \right]^{-1} 
&&
\text{Leontief economy}
\label{ltuni}
\end{align}
For the case of a Cobb-Douglas economy, where $\gamma=0$, we first take the log and let $\gamma \to 0$ where L'H\^{o}spital's rule is applicable since $\sum_{i=0}^n a_{ij}= 1$ for $j=1,\cdots,n$.
\begin{align}
\ln \pi_{j} z_{j} = \frac{\ln \left( \sum_{i=0}^n a_{ij} (\pi_{i})^\gamma \right)}{\gamma}
\to \frac{\sum_{i=0}^n a_{ij} (\pi_i)^\gamma \ln \pi_{i}}{\sum_{i=0}^n a_{ij} (\pi_i)^\gamma }
\to
\sum_{i=0}^n a_{ij} \ln \pi_{i}
\end{align}
Thus, (\ref{cesuni}) can be reduced in the following manner:
\begin{align}
\ln \bm{\pi} = \left( \bm{a}_0 \ln \pi_0  - \ln \bm{z} \right)
\left[ \mathbf{I} - \mathbf{A} \right]^{-1}
&&
\text{Cobb-Douglas economy}
\label{eqcd}
\end{align}
It is notable that the growth of the equilibrium price $\dif \ln \bm{\pi}=(\dif \ln \pi_1, \cdots, \dif \ln \pi_n)$ is a linear combination of the growth of sectoral productivity $\dif \ln \bm{z}=(\dif \ln z_1, \cdots, \dif \ln z_n)$ in the case of a Cobb-Douglas economy. 

Otherwise, the fixed point $\bm{\pi}$ given $\bm{z}$ can be searched for by using the simple recursive method applied to (\ref{eq}).
Since the unit cost function $\pi_j = (z_j)^{-1} c_j(\bm{\pi}; \pi_0)$ is monotonically increasing and strictly concave in $\bm{\pi}$, we know by \citet{kras} and \citet{kennan} that (\ref{eq}) is a contraction mapping that globally converges onto a unique fixed point, if it exists in $\mathbb{R}^{n}_{+}$.
Note that if $\pi_0=1$ and $(z_1,\cdots,z_n)=(1,\cdots,1)$, then $(\pi_1,\cdots,\pi_n)=(1, \cdots,1)$ is an equilibrium, which must be unique.
Moreover, note that obviously from (\ref{eqcd}), the existence of a positive fixed point $\bm{\pi} \in \mathbb{R}^n_{+}$ for any given $\bm{z} \in \mathbb{R}^n_{+}$ can be asserted for the case of a Cobb-Douglas economy.
Specifically, it is possible to show from (\ref{eqcd}) that:
\begin{align}
\left( \pi_1, \cdots, \pi_n \right)
= \left(
\prod_{i=1}^n \frac{e^{a_{01} \ell_{i1}\ln\pi_0}}{(z_i)^{\ell_{i1}}}
, \cdots ,
\prod_{i=1}^n \frac{e^{a_{0n} \ell_{in}\ln\pi_0}}{(z_i)^{\ell_{in}}}
\right)
> \left( 0, \cdots, 0 \right)
\end{align}
where $\ell_{ij}$ denotes the $ij$ element of the Leontief inverse $[\mathbf{I}-\mathbf{A}]^{-1}$.
Conversely, $\bm{\pi}$ can have negative elements or may not even exist in $\mathbb{R}^n$ for non-Cobb-Douglas economies. 
One may see this by replacing $\bm{z}$ with small (but positive) elements in (\ref{cesuni}) and (\ref{ltuni}).

\subsection{Viability of the equilibrium structure}
From another perspective, $c_j(\bm{\pi}; \pi_0)$ is homogeneous of degree one in $(\pi_0, \cdots, {\pi_n})$, so by Euler's homogeneous function theorem, it follows that:
\begin{align}
\pi_j = \sum_{i=1}^n (z_j)^{-1}\frac{\partial c_j}{\partial \pi_i} {\pi_i}  +(z_j)^{-1}\frac{\partial c_j}{\partial \pi_0} {\pi_0}
= \sum_{i=1}^n b_{ij} \pi_{i} + b_{0j} \pi_{0}
\end{align}
Here, by Shephard's lemma, $b_{ij}$ denotes the equilibrium physical input-output coefficient.
In matrix form, this is equivalent to:
\begin{align}
\bm{\pi}= \bm{\pi} \nabla \bm{c} \left< \bm{z} \right>^{-1} + \pi_0 \nabla_0 \bm{c} \left< \bm{z} \right>^{-1} = \bm{\pi} \mathbf{B} + \pi_0 \bm{b}_0 
\end{align}
Let us hereafter call $(\mathbf{B}, \bm{b}_0)$ the equilibrium structure (of an economy).
Note that if $\bm{\pi}$ exists in $\mathbb{R}^{n}_{+}$, while $\bm{b}_{0} \in \mathbb{R}^{n}_{+}$ (i.e., all sectors, upon production, physically utilize the primary factor), then $[\mathbf{I}-\mathbf{B}]$, where $\bm{\pi} \left[ \mathbf{I} - \mathbf{B} \right] =\pi_0 \bm{b}_0$, is said to satisfy the Hawkins-Simon (HS) condition \citep[Theorem 4.D.4][]{takayama, hsecta}.

The existence of a solution $\bm{y} \in \mathbb{R}^{n}_{+}$ for $[\mathbf{I}-\mathbf{B}] \bm{y} = \bm{d}$ given any $\bm{d} \in \mathbb{R}^{n}_{+}$ and the matrix $[\mathbf{I}-\mathbf{B}]$ satisfying the HS condition are two equivalent statements \citep[Theorem 4.D.1][]{takayama}.
Thus, a structure that $[\mathbf{I}-\mathbf{B}]$ satisfies the HS condition is said to be \textit{viable}.
Conversely, for an \textit{unviable} structure (that $[\mathbf{I}-\mathbf{B}]$ does not satisfy the HS condition), no positive production schedule $\bm{y} \in \mathbb{R}^{n}_{+}$ can be possible for fulfilling any positive final demand $\bm{d} \in \mathbb{R}^{n}_{+}$.
For a Cobb-Douglas economy, we can assert that the equilibrium structure is always viable, since we know from (\ref{eqcd}) that it is always the case that $\bm{\pi} \in \mathbb{R}^n_+$.
Otherwise, $\bm{\pi}$ may have negative elements, in which case the equilibrium structure must be unviable. 
An unviable equilibrium structure may never appear during the
recursive process, however, if the equilibrium price search is such
that it is installed in the recursive process of (\ref{eq}); instead,
the recursive process will not be convergent since (\ref{eq}) maps
into an open set $\mathbb{R}^n_+$. 

Last, let us specify below the structural transformation (as the
physical input-output coefficient matrix $\mathbf{B}$) and network
transformation (as the cost-share structure or the monetary
input-output coefficient matrix $\mathbf{S}$) given $\bm{z}$ in a uniform CES economy.
Since an element of the gradient of the CES aggregator is:
\begin{align}
\frac{\partial c_j}{\partial \pi_i}= a_{ij}(z_j)^{1-\gamma_j} \left( \frac{\pi_j}{\pi_i} \right)^{1-\gamma_j}
\end{align}
the gradient of the uniform CES aggregator can be written as follows:
\begin{align}
\nabla \bm{c} = \left< \bm{\pi} \right>^{\gamma-1}\mathbf{A}\left< \bm{\pi} \right>^{1-\gamma} \left< \bm{z} \right>^{1-\gamma}
\end{align}
Thus, below are the transformed structure and networks, where $\bm{\pi}$ is given by (\ref{cesuni}):
\begin{align}
\mathbf{B} 
=\left< \bm{\pi} \right>^{\gamma-1}\mathbf{A}\left< \bm{\pi} \right>^{1-\gamma} \left< \bm{z} \right>^{-\gamma},
&&
\mathbf{S} 
=\left< \bm{\pi} \right>^{\gamma}\mathbf{A}\left< \bm{\pi} \right>^{-\gamma} \left< \bm{z} \right>^{-\gamma}
\end{align}
Observe that $\mathbf{S}=\mathbf{A}$ in a Cobb-Douglas economy ($\gamma=0$) and $\mathbf{B}=\mathbf{A}\left<1/\bm{z} \right>$ in a Leontief economy ($\gamma=1$).

\section{Estimation}
Let us start by taking the log of (\ref{sij}) and indexing observations by $t=1,\cdots,T$, while omitting the sectoral index ($j$).
The cross-sectional dimension remains, i.e., $i=0,\cdots,n$.
Here, we substitute $p$ for $\pi$ to emphasize that they are observed
data, and $\zeta$ for $z$ to emphasize that they are parameters
subject to estimation.  
For the response variable, we use the factor share $a_{it}$ available as the input-output coefficient.
\begin{align}
\ln a_{it} = \ln \alpha_i - \gamma \ln (\zeta_t p_t) + \gamma \ln p_{it} + \epsilon_{it}
\label{reg}
\end{align}
Note that the error terms $\epsilon_{it}$ are assumed to be iid normally distributed with mean zero.
The multi-factor CES elasticity in which we are interested has been extensively studied in the Armington elasticity literature.
\cite{cje02} and \cite{saito} apply between estimation, a typical strategy for the two-input case, to estimate the elasticity of substitution between products from different countries.
Between estimation eliminates time-specific effects while saving the individual-specific effects such as the share parameter $\alpha_{i}$. 
For a two-factor case, the share parameter is usually subject to estimation. 
However, for a multi-factor case, the constraint that $\sum_{i=0}^n \alpha_{i} = 1$ can hardly be met. 
Moreover, we know in advance that $\alpha_i = a_{i}$ for the year that the model is standardized.
Hence, we opt to apply within (FE) estimation in this study.

Below we restate (\ref{reg}) using time dummy variables such that $\gamma$ and $\zeta_t p_t$ can be estimated from $p_{it}$ and $a_{it}$ via FE panel regression:
\begin{align}
Y_{it} = \mu_1 + (\mu_2 - \mu_1) D_2 + \cdots + ( \mu_T - \mu_1) D_T + \gamma X_{it} + \ln \alpha_{i} + \epsilon_{it}
\label{panel}
\end{align}
where $Y_{it}=\ln a_{it}$, $X_{it} = \ln p_{it}$ and $D_k$ for $k=2, \cdots, T$ denotes a dummy variable that equals $1$ if $k=t$ and $0$ otherwise.
For $t=1$, $D_2=\cdots=D_T=0$ by definition, so we know that $\mu_t=-\gamma \ln (\zeta_t p_t)$ for $t=1,\cdots,T$.
The estimable coefficients for (\ref{panel}) via FE, therefore, indicate that:
\begin{align}
\mu_t  - \mu_1 = - \gamma \left( \ln{\zeta_t p_t} - \ln{\zeta_1 p_1} \right)
&&
t = 2, \cdots, T
\end{align} 
We may thus evaluate the productivity growth at $t$, based on $t=1$, by the following formula:
\begin{align}
\ln \zeta_t/\zeta_1 = -(\mu_t-\mu_1)/\gamma - \ln p_t/p_1
\end{align} 

We face the concern that regression (\ref{reg}) suffers from an endogeneity problem.
The response variable, i.e., the demand for the $i$th factor of production by the $j$th sector, may well affect the price of the $i$th factor via the supply function.
Because of such reverse causality, the explanatory variable, i.e., the price of the $i$th commodity, becomes correlated with the error term that corresponds to the demand shock for the $i$th factor of production by the $j$th sector.
To remedy this problem, we apply total factor productivity (TFP) to
instrument prices. 
The \cite{jip} database provides sectoral TFP growth (in terms of the T\"{o}rnqvist index) as well as the aggregated macro-TFP growth, for each year interval.
It is generally assumed that TFP is unlikely to be correlated with the demand shock \citep{eslava, foster}.
In our case, the $i$th sector's TFP to produce the $i$th commodity is unlikely to be correlated with the $j$th sector's demand shock for the $i$th commodity. 
Hence, TFP appears to be suitable as an instrument for our explanatory variable.

On the other hand, the price of the primary factor $i=0$, can be nonresponsive to sectoral demand shocks.
The primary factor consists of labor and capital services, while their prices, i.e., wages and interest rates, are not purely dependent on the market mechanism but rather subject to government regulations and natural depreciations.
Moreover, it is conceivable that the demand shock for the primary factor by one sector has little influence on the prices of its factors, labor and capital, if not on their quantitative ratios demanded by the sector.
Thus, we apply three exogenous variables as instruments for $X_{0t}$, namely, 1) the macro TFP, 2) the macro wage rate, and 3) the macro interest rate, which are available in time series in the \citet{jip} database.
Specifically, we will be examining three instrumental variables in the FE IV regression of (\ref{panel}), namely, $v^a_{it}$, $v^b_{it}$, and $v^c_{it}$, all of which include the sectoral TFP at $t$, for $i=1,\cdots,n$, and where, $v^a_{0t}=\text{macro TFP at } t$, $v^b_{0t}=\text{macro wage rate at } t$, and $v^c_{0t}=\text{macro interest rate at } t$.

The results are summarized in Table \ref{tab:long}.
The first column (LS FE) reports the least squares fixed effects estimation results, without instrumenting for the explanatory variable.
The second column (IV FE) reports the instrumental variable fixed effects estimation results, using the IVs reported in the last column.
In all cases, overidentification tests are not rejected, so we are satisfied with the IVs we applied.
Furthermore, first-stage F values are large enough that we are satisfied with the strength of the IVs we applied.
Interestingly, the estimates for the elasticity of substitution $\hat{\sigma}=1-\hat{\gamma}$ are larger when IVs are applied.
For later study of the aggregate fluctuations, we select from the elasticity estimates based on the endogeneity test results. 
Specifically,  we use the LS FE estimates for sector ids 6, 12, 27, 52, 62, 70, 71, 81, 88 and, hence, IV FE estimates for the rest of the sectors.
Finally, we note that simple mean of the estimated (accepted) elasticity of substitution is $\bar{\hat{\sigma}}=1.54$.\footnote{
The uniform CES elasticity for the US production economy estimated by
\cite{atalay} using military spending as an IV is reportedly
approximately $-0.1$ with zero (Leontief) being unable to be rejected.}

\begin{center}
\begin{ThreePartTable}
\begin{TableNotes}
\footnotesize
\item[Notes:] For sector classifications (ids) see Table \ref{tab2}.
The household sector is {id} = 101.
Values in parentheses indicate p-values.
\item[*1] First-stage (Cragg-Donald Wald) F statistic for 2SLS FE estimation.
The rule of thumb to reject the hypothesis that the explanatory variable is only weakly correlated with the instrument is for this to exceed 10.
\item[*2] Overidentification test by Sargan statistic.
Rejection of the null indicates that the instruments are correlated with the residuals.
\item[*3] Endogeneity test by Davidson-MacKinnon F statistic.
Rejection of the null indicates that the instrumental variables fixed effects estimator should be employed.
\item[*4] Instrumental variables applied, where a, b, and c, indicate
  ${v^a}$, ${v^b}$, and ${v^c}$, respectively, and la, lb, and lc,
  indicate $\ln {v^a}$, $\ln {v^b}$, and $\ln {v^c}$, respectively.
\end{TableNotes}
\newcolumntype{.}{D{.}{.}{3}}
\newcolumntype{i}{D{.}{}{0}}
\begin{longtable}{r@{~~~}.@{~~~} . .@{~~~} .@{~} i .@{} .@{~~~~} .@{} . c}
\caption{Estimation of the elasticity of substitution for all 100 sectors.} \label{tab:long} \\
\hline\noalign{\smallskip}	
& \multicolumn{2}{c}{LS FE} & \multicolumn{8}{c}{IV FE} \\
\cmidrule(r){2-3}\cmidrule(r){4-11}
\multicolumn{1}{c}{id} &\multicolumn{1}{c}{$\hat{\sigma}$ } &\multicolumn{1}{c}{s.e.} &\multicolumn{1}{c}{$\hat{\sigma}$} &\multicolumn{1}{c}{s.e.} &\multicolumn{1}{c}{1st F\tnote{*1}} &\multicolumn{2}{c}{Overid.\tnote{*2}} &\multicolumn{2}{c}{Endog.\tnote{*3}}  &\multicolumn{1}{c}{IVs\tnote{*4}} \\ \hline\noalign{\smallskip}
\endfirsthead
\multicolumn{11}{c}%
{{\tablename\ \thetable{} -- continued from previous page}} \\
\hline\noalign{\smallskip}	
& \multicolumn{2}{c}{LS FE} & \multicolumn{8}{c}{IV FE} \\
\cmidrule(r){2-3}\cmidrule(r){4-11}
\multicolumn{1}{c}{id} &\multicolumn{1}{c}{$\hat{\sigma}$ } &\multicolumn{1}{c}{s.e.} &\multicolumn{1}{c}{$\hat{\sigma}$} &\multicolumn{1}{c}{s.e.} &\multicolumn{1}{c}{1st F\tnote{*1}} &\multicolumn{2}{c}{Overid.\tnote{*2}} &\multicolumn{2}{c}{Endog.\tnote{*3}}  &\multicolumn{1}{c}{IVs\tnote{*4}} \\ \hline\noalign{\smallskip}
\endhead
\hline
\endfoot
\hline 
\insertTableNotes
\endlastfoot
1	&	1.119		&	0.082		&	3.540		&	0.317		&	106		&	0.11		&	(0.735)		&	92.07		&	(0.000)		&	la, c		\\
2	&	1.422		&	0.093		&	2.739		&	0.273		&	145		&	0.44		&	(0.507)		&	29.21		&	(0.000)		&	a, c		\\
3	&	0.994		&	0.065		&	2.850		&	0.262		&	94		&	0.79		&	(0.373)		&	76.80		&	(0.000)		&	lc, c		\\
4	&	1.178		&	0.065		&	1.947		&	0.154		&	228		&	1.42		&	(0.234)		&	32.66		&	(0.000)		&	lb, c		\\
5	&	0.782		&	0.054		&	2.104		&	0.150		&	197		&	0.99		&	(0.319)		&	121.19		&	(0.000)		&	la, c		\\
6	&	1.167		&	0.071		&	1.288		&	0.202		&	139		&	0.72		&	(0.397)		&	0.41		&	(0.523)		&	la, c		\\
7	&	0.936		&	0.062		&	1.698		&	0.192		&	124		&	0.18		&	(0.668)		&	19.12		&	(0.000)		&	la, c		\\
8	&	0.724		&	0.062		&	1.118		&	0.137		&	260		&	0.02		&	(0.884)		&	10.63		&	(0.001)		&	la, c		\\
9	&	0.553		&	0.070		&	1.384		&	0.183		&	187		&	0.15		&	(0.696)		&	26.18		&	(0.000)		&	la, c		\\
10	&	0.962		&	0.076		&	1.519		&	0.275		&	86		&	0.10		&	(0.752)		&	4.56		&	(0.033)		&	a, c		\\
11	&	0.671		&	0.076		&	2.145		&	0.261		&	111		&	0.10		&	(0.752)		&	42.06		&	(0.000)		&	la, c		\\
12	&	1.052		&	0.093		&	1.312		&	0.264		&	139		&	0.10		&	(0.754)		&	1.11		&	(0.292)		&	la, c		\\
13	&	0.389		&	0.064		&	1.190		&	0.177		&	163		&	0.87		&	(0.351)		&	25.61		&	(0.000)		&	la, c		\\
14	&	0.805		&	0.048		&	1.241		&	0.115		&	217		&	1.45		&	(0.229)		&	18.33		&	(0.000)		&	a, c		\\
15	&	0.608		&	0.048		&	1.402		&	0.114		&	250		&	0.65		&	(0.421)		&	68.15		&	(0.000)		&	la, c		\\
16	&	0.153		&	0.051		&	0.898		&	0.116		&	179		&	3.60		&	(0.166)		&	58.22		&	(0.000)		&	la, lc	\\
17	&	0.840		&	0.059		&	2.723		&	0.184		&	183		&	1.84		&	(0.174)		&	186.29		&	(0.000)		&	la, lc		\\
18	&	0.723		&	0.047		&	1.515		&	0.133		&	160		&	0.69		&	(0.407)		&	46.70		&	(0.000)		&	la, lc		\\
19	&	0.479		&	0.049		&	2.605		&	0.233		&	94		&	1.55		&	(0.214)		&	176.49		&	(0.000)		&	la, c		\\
20	&	0.880		&	0.047		&	1.888		&	0.136		&	176		&	0.00		&	(0.998)		&	79.31		&	(0.000)		&	la, c		\\
21	&	0.637		&	0.047		&	0.978		&	0.098		&	303		&	0.93		&	(0.335)		&	16.09		&	(0.000)		&	la, c		\\
22	&	0.533		&	0.049		&	0.781		&	0.102		&	307		&	0.00		&	(0.997)		&	7.90		&	(0.005)		&	la, c		\\
23	&	-0.111		&	0.047		&	1.946		&	0.243		&	78		&	1.49		&	(0.223)		&	150.55		&	(0.000)		&	la, c		\\
24	&	0.502		&	0.048		&	1.765		&	0.211		&	74		&	0.16		&	(0.687)		&	51.81		&	(0.000)		&	la, c		\\
25	&	0.605		&	0.038		&	0.908		&	0.069		&	455		&	1.73		&	(0.188)		&	28.83		&	(0.000)		&	la, c		\\
26	&	0.855		&	0.046		&	1.644		&	0.105		&	285		&	1.83		&	(0.176)		&	82.84		&	(0.000)		&	la, c		\\
27	&	0.158		&	0.050		&	0.200		&	0.104		&	302		&	0.04		&	(0.844)		&	0.21		&	(0.643)		&	la, c		\\
28	&	0.585		&	0.042		&	1.446		&	0.105		&	241		&	0.13		&	(0.716)		&	101.26		&	(0.000)		&	la, c		\\
29	&	0.624		&	0.054		&	3.700		&	0.645		&	19		&	0.02		&	(0.901)		&	60.56		&	(0.000)		&	b, lc		\\
30	&	0.510		&	0.052		&	1.295		&	0.192		&	88		&	0.00		&	(0.961)		&	20.24		&	(0.000)		&	b, lc		\\
31	&	0.641		&	0.044		&	2.076		&	0.185		&	94		&	0.33		&	(0.565)		&	101.52		&	(0.000)		&	b, lc		\\
32	&	0.890		&	0.041		&	1.209		&	0.106		&	185		&	1.96		&	(0.162)		&	10.96		&	(0.001)		&	b, lc		\\
33	&	0.414		&	0.047		&	1.215		&	0.140		&	148		&	0.03		&	(0.854)		&	43.16		&	(0.000)		&	b, a		\\
34	&	0.325		&	0.044		&	0.978		&	0.102		&	260		&	0.66		&	(0.417)		&	57.24		&	(0.000)		&	b, a		\\
35	&	0.500		&	0.046		&	1.095		&	0.115		&	203		&	0.92		&	(0.337)		&	34.54		&	(0.000)		&	b, a		\\
36	&	0.615		&	0.047		&	1.054		&	0.113		&	223		&	0.06		&	(0.808)		&	19.04		&	(0.000)		&	b, a		\\
37	&	0.434		&	0.046		&	0.787		&	0.084		&	450		&	0.96		&	(0.327)		&	26.22		&	(0.000)		&	b, la		\\
38	&	0.410		&	0.049		&	0.907		&	0.106		&	286		&	1.05		&	(0.306)		&	29.97		&	(0.000)		&	lc, c		\\
39	&	0.656		&	0.047		&	1.480		&	0.108		&	272		&	0.12		&	(0.728)		&	84.98		&	(0.000)		&	a, c		\\
40	&	0.410		&	0.045		&	0.876		&	0.087		&	382		&	0.65		&	(0.421)		&	41.80		&	(0.000)		&	a, c		\\
41	&	0.490		&	0.041		&	0.975		&	0.066		&	698		&	1.13		&	(0.289)		&	98.99		&	(0.000)		&	a, c		\\
42	&	0.460		&	0.045		&	1.022		&	0.099		&	285		&	1.48		&	(0.224)		&	44.20		&	(0.000)		&	a, c		\\
43	&	0.746		&	0.039		&	1.400		&	0.085		&	318		&	0.02		&	(0.876)		&	89.63		&	(0.000)		&	la, c		\\
44	&	0.199		&	0.045		&	0.689		&	0.094		&	318		&	0.23		&	(0.634)		&	37.50		&	(0.000)		&	a, c		\\
45	&	0.777		&	0.043		&	1.482		&	0.081		&	473		&	0.14		&	(0.709)		&	124.74		&	(0.000)		&	la, c		\\
46	&	0.725		&	0.044		&	1.192		&	0.083		&	414		&	0.03		&	(0.859)		&	47.79		&	(0.000)		&	la, c		\\
47	&	0.438		&	0.044		&	1.050		&	0.084		&	432		&	0.54		&	(0.464)		&	83.15		&	(0.000)		&	a, c		\\
48	&	0.199		&	0.042		&	0.938		&	0.090		&	343		&	0.00		&	(0.947)		&	103.70		&	(0.000)		&	a, c		\\
49	&	0.698		&	0.045		&	1.182		&	0.111		&	205		&	0.48		&	(0.489)		&	24.00		&	(0.000)		&	a, c		\\
50	&	0.467		&	0.048		&	0.807		&	0.109		&	244		&	1.24		&	(0.266)		&	12.39		&	(0.000)		&	a, c		\\
51	&	0.471		&	0.039		&	1.057		&	0.092		&	252		&	0.10		&	(0.758)		&	56.61		&	(0.000)		&	a, c		\\
52	&	0.574		&	0.061		&	0.682		&	0.142		&	223		&	0.65		&	(0.420)		&	0.71		&	(0.399)		&	a, c		\\
53	&	0.786		&	0.062		&	1.685		&	0.154		&	213		&	0.04		&	(0.848)		&	45.58		&	(0.000)		&	a, c		\\
54	&	0.797		&	0.051		&	1.535		&	0.139		&	175		&	0.05		&	(0.831)		&	36.42		&	(0.000)		&	a, c		\\
55	&	0.794		&	0.048		&	1.410		&	0.135		&	158		&	0.25		&	(0.619)		&	25.89		&	(0.000)		&	a, c		\\
56	&	0.749		&	0.052		&	1.126		&	0.152		&	137		&	0.18		&	(0.675)		&	7.17		&	(0.007)		&	a, c		\\
57	&	0.829		&	0.066		&	1.628		&	0.172		&	185		&	0.03		&	(0.870)		&	27.69		&	(0.000)		&	a, c		\\
58	&	0.122		&	0.054		&	0.338		&	0.114		&	281		&	0.66		&	(0.416)		&	4.60		&	(0.032)		&	a, c		\\
59	&	0.319		&	0.041		&	0.704		&	0.087		&	300		&	1.54		&	(0.215)		&	26.23		&	(0.000)		&	a, c		\\
60	&	0.134		&	0.057		&	2.522		&	0.246		&	111		&	2.67		&	(0.102)		&	196.47		&	(0.000)		&	la, c		\\
61	&	0.160		&	0.047		&	2.548		&	0.218		&	118		&	0.30		&	(0.586)		&	308.57		&	(0.000)		&	la, c		\\
62	&	1.114		&	0.070		&	1.011		&	0.218		&	113		&	0.13		&	(0.723)		&	0.25		&	(0.618)		&	a, c		\\
63	&	1.222		&	0.067		&	3.412		&	0.249		&	120		&	0.13		&	(0.722)		&	135.59		&	(0.000)		&	lc, a		\\
64	&	1.028		&	0.111		&	2.768		&	0.476		&	65		&	1.80		&	(0.180)		&	15.97		&	(0.000)		&	lc, a		\\
65	&	0.863		&	0.053		&	1.696		&	0.130		&	233		&	0.21		&	(0.651)		&	57.10		&	(0.000)		&	a, c		\\
66	&	1.004		&	0.050		&	0.758		&	0.121		&	210		&	0.87		&	(0.350)		&	5.06		&	(0.024)		&	lc, a		\\
67	&	1.206		&	0.054		&	1.591		&	0.161		&	129		&	0.96		&	(0.326)		&	6.61		&	(0.010)		&	la, c		\\
68	&	1.112		&	0.062		&	1.477		&	0.123		&	345		&	0.94		&	(0.334)		&	12.11		&	(0.001)		&	lc, a		\\
69	&	1.164		&	0.078		&	2.641		&	0.160		&	383		&	0.86		&	(0.353)		&	137.82		&	(0.000)		&	lb, a		\\
70	&	0.920		&	0.047		&	1.012		&	0.101		&	275		&	0.48		&	(0.488)		&	1.07		&	(0.302)		&	a, c		\\
71	&	0.960		&	0.051		&	0.979		&	0.106		&	296		&	0.69		&	(0.407)		&	0.04		&	(0.838)		&	a, c		\\
72	&	1.048		&	0.050		&	1.864		&	0.124		&	223		&	0.14		&	(0.713)		&	60.48		&	(0.000)		&	a, c		\\
73	&	1.513		&	0.071		&	3.985		&	0.261		&	135		&	0.10		&	(0.758)		&	162.65		&	(0.000)		&	lc, c		\\
74	&	0.890		&	0.061		&	1.556		&	0.145		&	231		&	1.03		&	(0.311)		&	27.46		&	(0.000)		&	la, c		\\
75	&	0.703		&	0.060		&	1.375		&	0.134		&	260		&	0.99		&	(0.319)		&	33.47		&	(0.000)		&	lc, c		\\
76	&	0.887		&	0.059		&	1.644		&	0.133		&	265		&	0.55		&	(0.460)		&	44.19		&	(0.000)		&	lb, c		\\
77	&	0.904		&	0.049		&	1.448		&	0.111		&	257		&	0.97		&	(0.326)		&	32.11		&	(0.000)		&	lb, c		\\
78	&	1.152		&	0.077		&	1.514		&	0.154		&	340		&	0.95		&	(0.331)		&	7.52		&	(0.006)		&	lb, c		\\
79	&	0.647		&	0.058		&	0.926		&	0.133		&	239		&	0.27		&	(0.603)		&	5.49		&	(0.019)		&	b, c		\\
80	&	0.617		&	0.056		&	1.223		&	0.125		&	270		&	0.14		&	(0.712)		&	31.68		&	(0.000)		&	b, c		\\
81	&	0.601		&	0.052		&	0.649		&	0.110		&	285		&	0.30		&	(0.586)		&	0.24		&	(0.625)		&	lb, a		\\
82	&	0.702		&	0.058		&	1.337		&	0.129		&	272		&	0.00		&	(0.995)		&	32.48		&	(0.000)		&	b, c		\\
83	&	1.057		&	0.061		&	2.014		&	0.143		&	255		&	0.09		&	(0.763)		&	62.94		&	(0.000)		&	b, c		\\
84	&	1.149		&	0.153		&	2.067		&	0.422		&	147		&	0.04		&	(0.836)		&	5.57		&	(0.018)		&	lb, a		\\
85	&	1.147		&	0.059		&	1.989		&	0.130		&	285		&	0.73		&	(0.393)		&	59.29		&	(0.000)		&	lc, c		\\
86	&	0.864		&	0.063		&	1.618		&	0.137		&	290		&	1.06		&	(0.304)		&	41.69		&	(0.000)		&	lb, a		\\
87	&	1.136		&	0.073		&	1.891		&	0.167		&	249		&	1.04		&	(0.309)		&	26.77		&	(0.000)		&	la, c		\\
88	&	0.667		&	0.049		&	0.755		&	0.095		&	365		&	0.70		&	(0.404)		&	1.15		&	(0.283)		&	a, c		\\
89	&	0.738		&	0.056		&	1.578		&	0.126		&	282		&	1.18		&	(0.277)		&	63.45		&	(0.000)		&	a, c		\\
90	&	0.990		&	0.050		&	1.719		&	0.138		&	170		&	1.02		&	(0.314)		&	36.10		&	(0.000)		&	a, c		\\
91	&	0.831		&	0.056		&	1.578		&	0.129		&	264		&	0.00		&	(0.999)		&	46.26		&	(0.000)		&	a, c		\\
92	&	1.241		&	0.054		&	2.084		&	0.134		&	227		&	0.56		&	(0.456)		&	54.39		&	(0.000)		&	la, c		\\
93	&	1.073		&	0.076		&	1.616		&	0.192		&	191		&	0.17		&	(0.676)		&	9.78		&	(0.002)		&	a, c		\\
94	&	0.970		&	0.053		&	1.495		&	0.111		&	311		&	0.28		&	(0.596)		&	30.94		&	(0.000)		&	la, c		\\
95	&	0.253		&	0.076		&	1.757		&	0.223		&	122		&	0.06		&	(0.804)		&	68.04		&	(0.000)		&	la, c		\\
96	&	0.571		&	0.054		&	1.542		&	0.122		&	290		&	0.38		&	(0.536)		&	94.91		&	(0.000)		&	la, c		\\
97	&	0.887		&	0.054		&	1.485		&	0.124		&	247		&	0.07		&	(0.799)		&	30.79		&	(0.000)		&	la, c		\\
98	&	0.559		&	0.067		&	1.160		&	0.159		&	228		&	2.61		&	(0.106)		&	18.15		&	(0.000)		&	la, c		\\
99	&	0.771		&	0.063		&	1.479		&	0.144		&	253		&	0.03		&	(0.865)		&	32.33		&	(0.000)		&	la, c		\\
100	&	0.289		&	0.087		&	2.412		&	0.290		&	132		&	0.25		&	(0.616)		&	78.53		&	(0.000)		&	c, lc		\\
101	&	0.997		&	0.043		&	1.198		&	0.091		&	389		&	0.56		&	(0.456)		&	21.20		&	(0.000)		&	a, a		\\

\end{longtable}
\end{ThreePartTable}
\end{center}

\section{Aggregate Fluctuations}
\subsection{Representative Household}
Let us now consider a representative household that maximizes the following CES utility:
\begin{align}
H(\bm{h})
= \left( (\mu_1)^{\frac{1}{1-\kappa}} (h_1)^{\frac{\kappa}{1-\kappa}} + \cdots +  (\mu_n)^{\frac{1}{1-\kappa}} (h_n)^{\frac{\kappa}{1-\kappa}} \right)^{\frac{1-\kappa}{\kappa}} 
\end{align}
The household determines the consumption schedule $\bm{h}=(h_1, \cdots, h_n)^\intercal$ given the budget constraint $W = \sum_{i=1}^n \pi_i h_i$ and prices of all goods $\bm{\pi}=(\pi_1, \cdots, \pi_n)$.
The source of the budget is the renumeration for the household's supply of the primary factor to the production sectors, so we know that $ W = \sum_{j=1}^n v_{j}$ (total value added, or GDP of the economy) is the representative household's (or national) income. 
The indirect utility of the household can then be specified as follows:
\begin{align}
H(\bm{h}(\bm{\pi}, W))
= \frac{W}{\left( \mu_1 (\pi_1)^{\kappa} + \cdots +  \mu_n (\pi_n)^{\kappa} \right)^{1/\kappa}}
=\frac{W}{\Pi(\bm{\pi})} 
\label{iu}
\end{align}
where $\Pi$, as defined as above, denotes the representative household's CES price index.
Note that $H=W$ at the baseline $(z_1, \cdots, z_n)=(1, \cdots, 1)$ where $\Pi =1$. 
Thus, $H$ is the utility (in terms of money) that the representative household can obtain from its income $W$ given the price change $\bm{\pi}$ (due to the productivity shock $\bm{z}$) while holding the primary input's price constant at $\pi_0 =1$.
In other words, $H$ is the real GDP if $W$ is the nominal GDP.

From another perspective, we note that the household's income $W$ can also be affected by the productivity shock. 
When there is a productivity gain in a production process, this process can either increase its output while holding all its inputs fixed or reduce the inputs while holding the output fixed.
In the former case, the national income $W$ (nominal GDP) remains at the baseline level, which equals real GDP in the previous year $\overline{H}$, and GDP growth ($\Delta \ln H = \ln H - \ln \overline{H}=-\ln \Pi(\bm{\pi})$) is fully accounted for. 
In the latter case, however, the national income can be reduced by as much as $W=\overline{H}\Pi(\bm{\pi})$, in which case we have $\Delta \ln H=0$ i.e., no GDP growth will be accounted for.

Of course, the reality must be in between the two extreme cases.
In this study, we conservatively evaluate national income (as nominal GDP) to the following extent:
\begin{align}
W = \overline{H} \left( \mu_1 (1/z_1)^\kappa + \cdots +\mu_n (1/z_n)^\kappa \right)^{1/\kappa}
= \overline{H}\Pi(1/\bm{z})
\end{align}
The real GDP under the equilibrium price, which equals the household's expenditure, can then be evaluated as follows:
\begin{align}
\ln H = \ln {W} - \ln {\Pi( \bm{\pi} )}
=\ln \overline{H}  
-\ln {\Pi( \bm{\pi} )} + \ln {\Pi(1/\bm{z})}
\label{lnH}
\end{align}
If we assume Cobb-Douglas utility ($\kappa \to 0$) and normalize the initial real GDP ($\overline{H}=1$), we have the following exposition:
\begin{align}
\ln H 
= -\ln \Pi(\bm{\pi}) +\Pi(1/\bm{z})
= -  \left(\ln \bm{\pi} + \ln \bm{z} \right) \bm{\mu}
\label{lnHcd}
\end{align}
where $\bm{\mu}=(\mu_1,\cdots,\mu_n)^\intercal$ denotes the column vector of expenditure share parameters.
The first identity indicates that the real GDP growth is the negative price index growth of the economy less the negative price index growth of a simple economy.\footnote{In a simple economy where there is no intermediate production, (\ref{eq}) is reduced as $\bm{\pi}=\bm{z}^{-1}$.}
Moreover, if we assume a Cobb-Douglas economy ($\gamma_j \to 0$), we arrive at the following:
\begin{align}
{\ln H} = - (\ln \bm{z} ) \left( \left[ \mathbf{I} - \mathbf{A} \right]^{-1} + \mathbf{I} \right) \bm{\mu} 
=
\sum_{j=1}^n \lambda_j \ln z_j
\end{align}
Note that $\lambda_j$ is the Domar weight \citep{hulten} in this particular case.

The parameters of the utility function are also subject to estimation.
By applying Roy's identity, i.e., $h_i = - \frac{\partial H}{\partial p_i}/\frac{\partial H}{\partial W}$, we have the following expansion for the household's expenditure share on the $i$th commodity:
\begin{align}
s_i =
\frac{\pi_i h_i}{\sum_{i=1}^n \pi_i h_i} = \frac{\mu_i (\pi_i)^\kappa}{\sum_{i=1}^n \mu_i (\pi_i)^\kappa}
= \mu_i \left( \frac{\Pi}{\pi_i} \right)^{-\kappa}
\label{ho}
\end{align}
where, $s_{i}$ denotes the expenditure share of the $i$th commodity for the representative household.
By taking the log of (\ref{ho}) and indexing observations by $t=1,\cdots,T$, we obtain the following regression equation where the parameter $\kappa$ can be estimated via FE.
\begin{align}
\ln m_{it} = \ln \mu_i - \kappa \ln (\Pi_t) + \kappa \ln p_{it} + \delta_{it}
\label{regho}
\end{align}
As is typical, the error term $\delta_{it}$ is assumed to be iid normally distributed with mean zero.
Here, we replace $\pi$ with $p$ to emphasize that they are observed data. 
For the response variable, we use the expenditure share $m_{it}$ of the final demand available from the input-output tables.
The cross-sectional dimension of the data for regression equation (\ref{regho}) is $i=1,\cdots,n$, whereas it is $i=0,1,\cdots,n$ for (\ref{reg}).
Thus, we apply sectoral TFP available for $t=1, \cdots, T$ from the \citet{jip} database as instruments to fix the endogeneity of the explanatory variable.
The estimation result for $\kappa$ using time dummy variables as in (\ref{panel}) (such that we may retrieve the estimates for $\Pi_t$) is presented in Table \ref{tab:long} (id = 101).

\subsection{Tail asymmetry and robustness}

\begin{figure}[t!]
\centering
\includegraphics[width=0.495\textwidth]{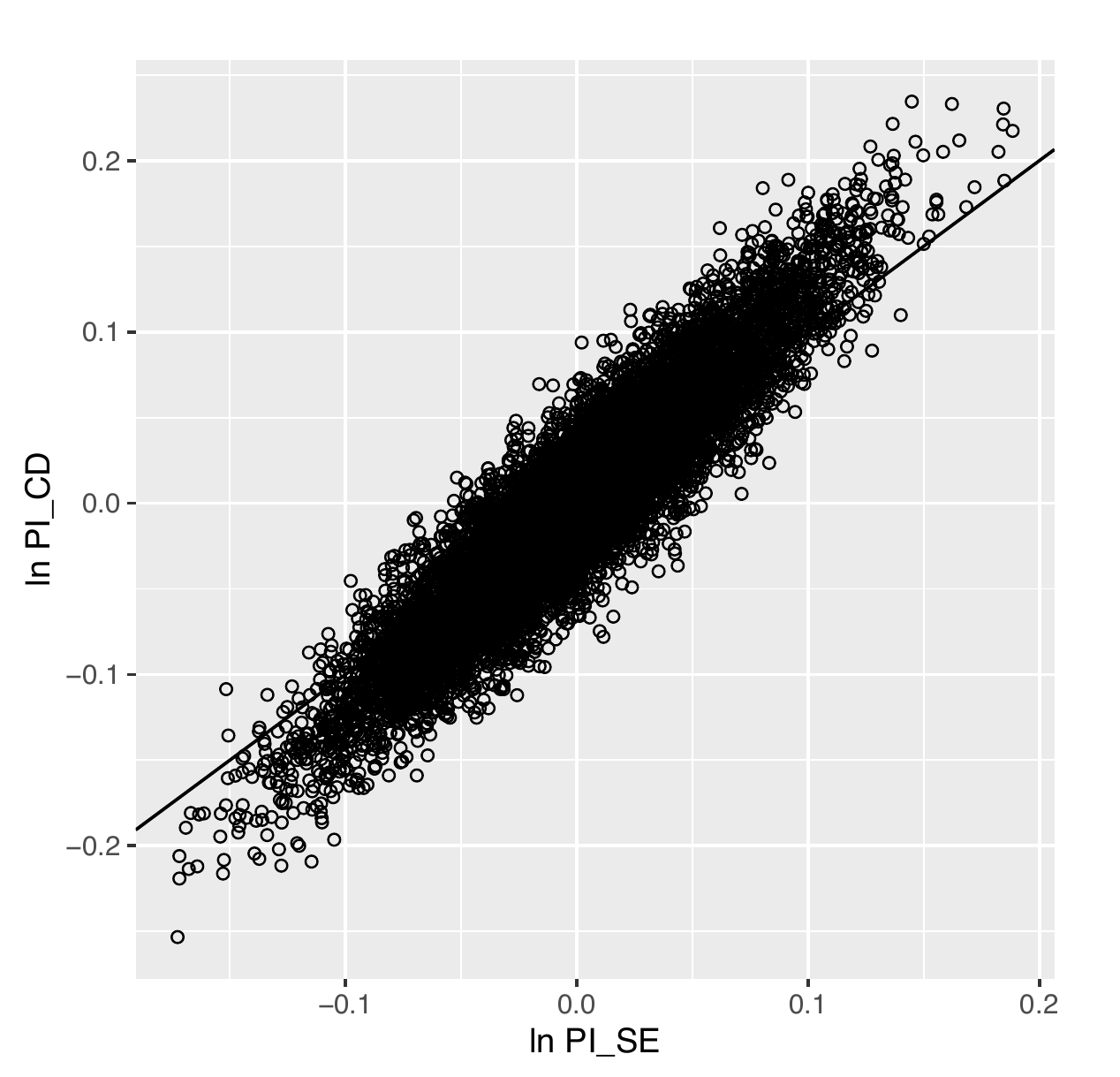}
\includegraphics[width=0.495\textwidth]{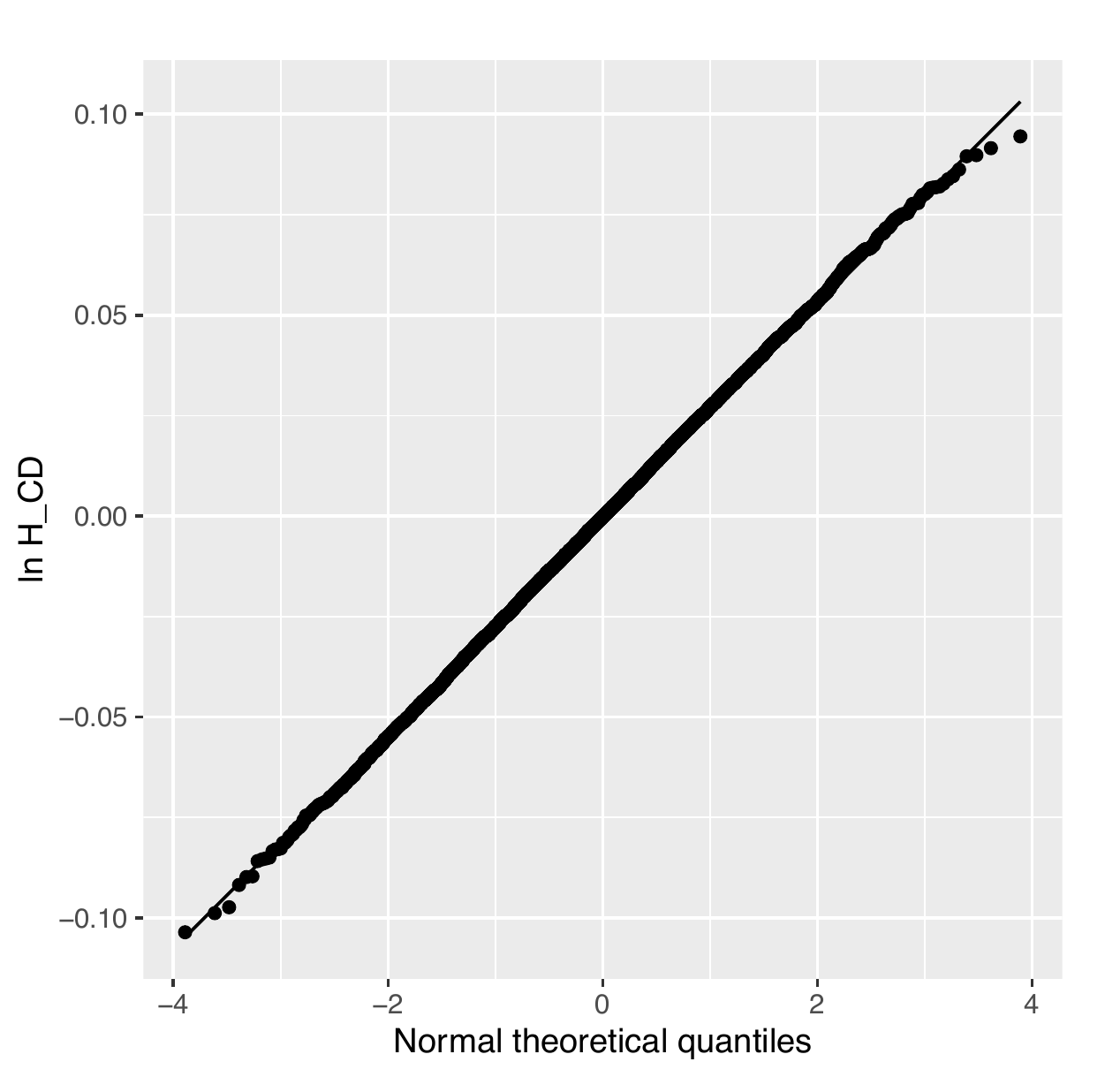}
\caption{
Left: Price index dispersion in Cobb-Douglas and Simple economies under Cobb-Douglas utility.
Dispersion is larger in the Cobb-Douglas than in the Simple economy, due to the power-law granularity of the Leontief inverse.
Right: QQ plot of the aggregate output fluctuations generated by the
linear Domar aggregator for the Cobb-Douglas economy and utility.
} \label{fig1}
\end{figure}

For a quantitative illustration, we study the distribution of aggregate output $\ln H$ when sectoral shocks $\ln \bm{z}$ are drawn from a normal distribution.
Specifically, we use 10,000 $\ln z_j \sim \mathcal{N}(0, 0.2)$ iid samples for $j=1,\cdots,n$, where the standard deviation (i.e., annual volatility of 20\%) is chosen with reference to the annual volatility of the estimated sectoral productivity growth $\ln \zeta_j$ (see \ref{app1}). 
Let us first examine the granularity of our baseline production networks (i.e., 2011 input-output linkages).
Below are both Cobb-Douglas price indices in terms of productivity shocks $\ln\bm{z}$ for the Cobb-Douglas and simple economies:
\begin{align}
\ln \Pi_{\text{CD}} = - (\ln \bm{z}) \left[\mathbf{I} - \mathbf{A} \right]^{-1} \bm{m},
&&
\ln \Pi_{\text{SE}} = - \left( \ln \bm{z} \right)\bm{m} 
\end{align}
Here, we set the share parameter $\bm{\mu}$ at the standard expenditure share of the final demand $\bm{m}=(m_1, \cdots, m_n)^\intercal$ for the year 2011.
Both indices must follow normal distributions because they are both linear with respect to the normal shocks $\ln \bm{z}$.  
The variances differ, however, and the left panel of Figure \ref{fig1} depicts the difference.
Observe the dilation of the variance in the Cobb-Douglas economy where the power-law granularity of the Leontief inverse causes the difference \citep{gabaixECTA, aceECTA}.

Replacing the equilibrium price of (\ref{lnH}) with the output of (\ref{eqe}) yields the following Domar aggregator, where exogenous productivity shocks ($\ln \bm{z}$) are aggregated into the growth of output ($\ln H$) for a CES economy.
\begin{align}
\ln H 
= - \ln \Pi\left( \mathcal{E}\left( \bm{z}; \pi_0=1 \right) \right) + \ln \Pi\left(  1/\bm{z} \right)
\end{align}
Note that this aggregator involves recursion in $\mathcal{E}$, as
specified by (\ref{eq}), for a CES economy with non-uniform
substitution elasticities. 
In this section, Cobb-Douglas utility is assumed for comparison with previous research, whence the Domar aggregator becomes:
\begin{align}
\ln H 
= - \left( \ln \mathcal{E}\left( \bm{z}\right) + \ln \bm{z} \right) \bm{m} 
&&
\text{CES economy}
\label{domarces}
\end{align}
Again, we set the share parameter $\bm{\mu}$ at the standard expenditure share of the final demand. 
In the case of the Leontief economy, the closed form is available from (\ref{ltuni}) as follows:
\begin{align}
\ln H 
= -  \left( \ln \left( \bm{a}_0 \left[ \left< \bm{z} \right> - \mathbf{A} \right]^{-1} \right) + \ln \bm{z} \right)\bm{m}
&&
\text{Leontief economy}
\label{domarlt}
\end{align}
The case for the Cobb-Douglas economy is also obtainable from (\ref{eqcd}) as follows:
\begin{align}
\ln H 
= - (\ln \bm{z}) \left( \left[ \mathbf{I} - \mathbf{A} \right]^{-1}  + \mathbf{I} \right)\bm{m}
&&
\text{Cobb-Douglas economy}
\label{domarcd}
\end{align}

\begin{figure}[t!]
\centering
\includegraphics[width=0.495\textwidth]{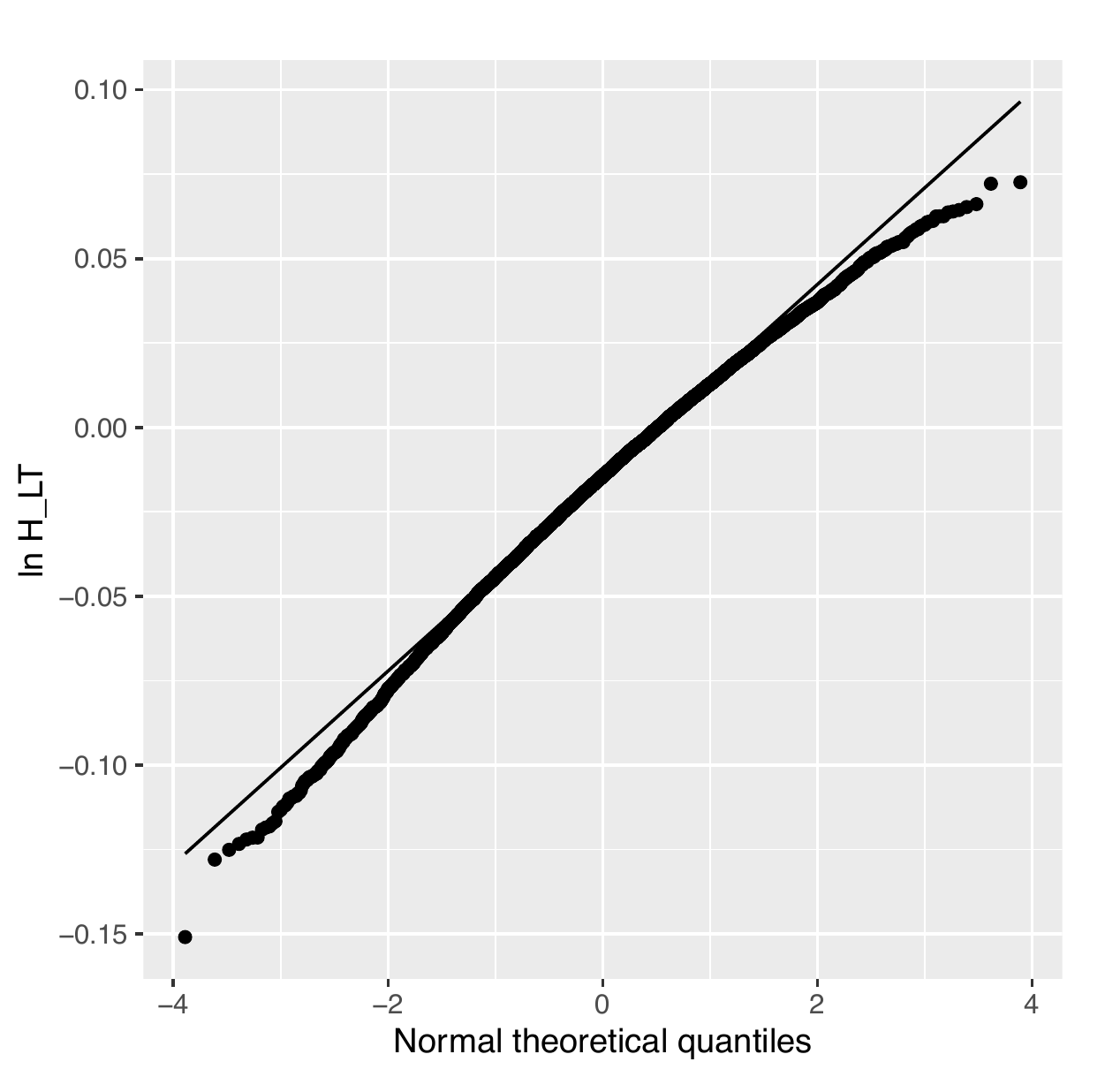}
\includegraphics[width=0.495\textwidth]{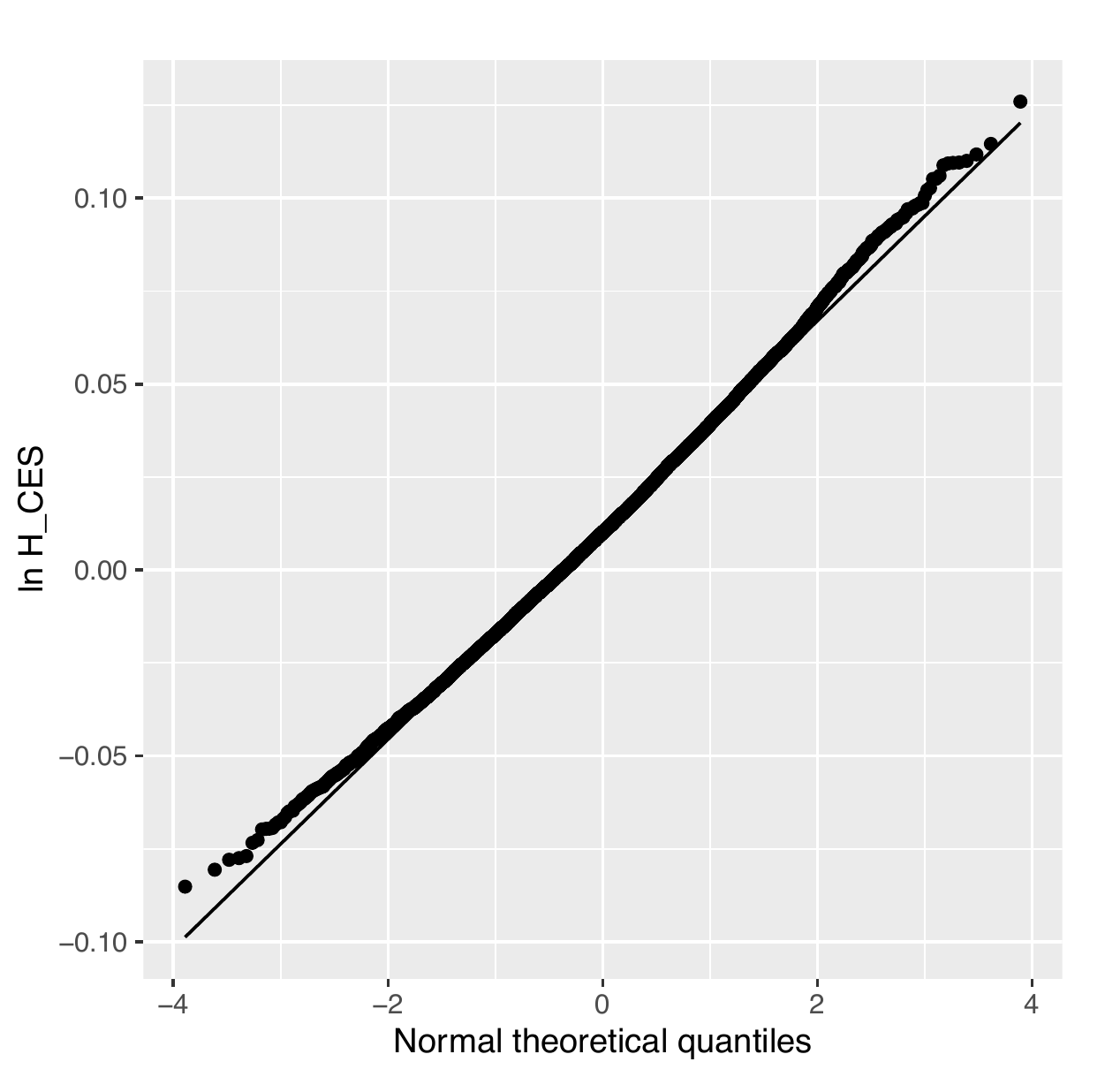}
\caption{
QQ plots of the aggregate output fluctuations generated by the Domar aggregators for Leontief (left) and elastic CES (right) economies under Cobb-Douglas utility.
} \label{fig2}
\end{figure}
As is obvious from the linearity of (\ref{domarcd}), the aggregate fluctuations must be normally distributed in the Cobb-Douglas economy. 
The resulting QQ plot is depicted in the right panel of Figure \ref{fig1}.
Note further that $\mathbb{E} [\ln H] = 0$ because of the linearity of (\ref{domarcd}) and $\mathbb{E}[\ln \bm{z}]=(0,\cdots,0)$; this is recognizable in the same figure.
That is, the expected economic growth is zero against zero-mean turbulences.
In other words, the unit elasticity of the Cobb-Douglas economy precisely absorbs the turbulences as if there were none to maintain zero expected growth.
Such \text{robustness} is absent in a Leontief economy with zero elasticity of substitution. 
The left panel of Figure \ref{fig2} illustrates the resulting QQ plot of the aggregate fluctuations generated by the same zero-mean normal shocks by way of (\ref{domarlt}).\footnote{Note that several sample normal shocks (with annual volatility of 20\%) made the equilibrium structure \textit{unviable} in the Leontief economy.
Such samples are excluded from the left panel of Figure \ref{fig2}.}
In this case, we observe negative expected output growth, i.e., $\overline{\ln H} =-1.57\%$, whose absolute value, in turn, can be interpreted as the robustness of the unit elasticity of substitution.\footnote{\citet{baqaee} also provide the foundation that an elastic (inelastic) economy implies positive (negative) expected output growth, based on the second-order approximation of nonlinear Domar aggregators.}
Moreover, we observe that normal shocks to the Leontief economy result in aggregate fluctuations with tail asymmetry similar to that depicted in the left panel of Figure \ref{fig0}.

\begin{figure}[t!]
\centering
\includegraphics[width=0.495\textwidth]{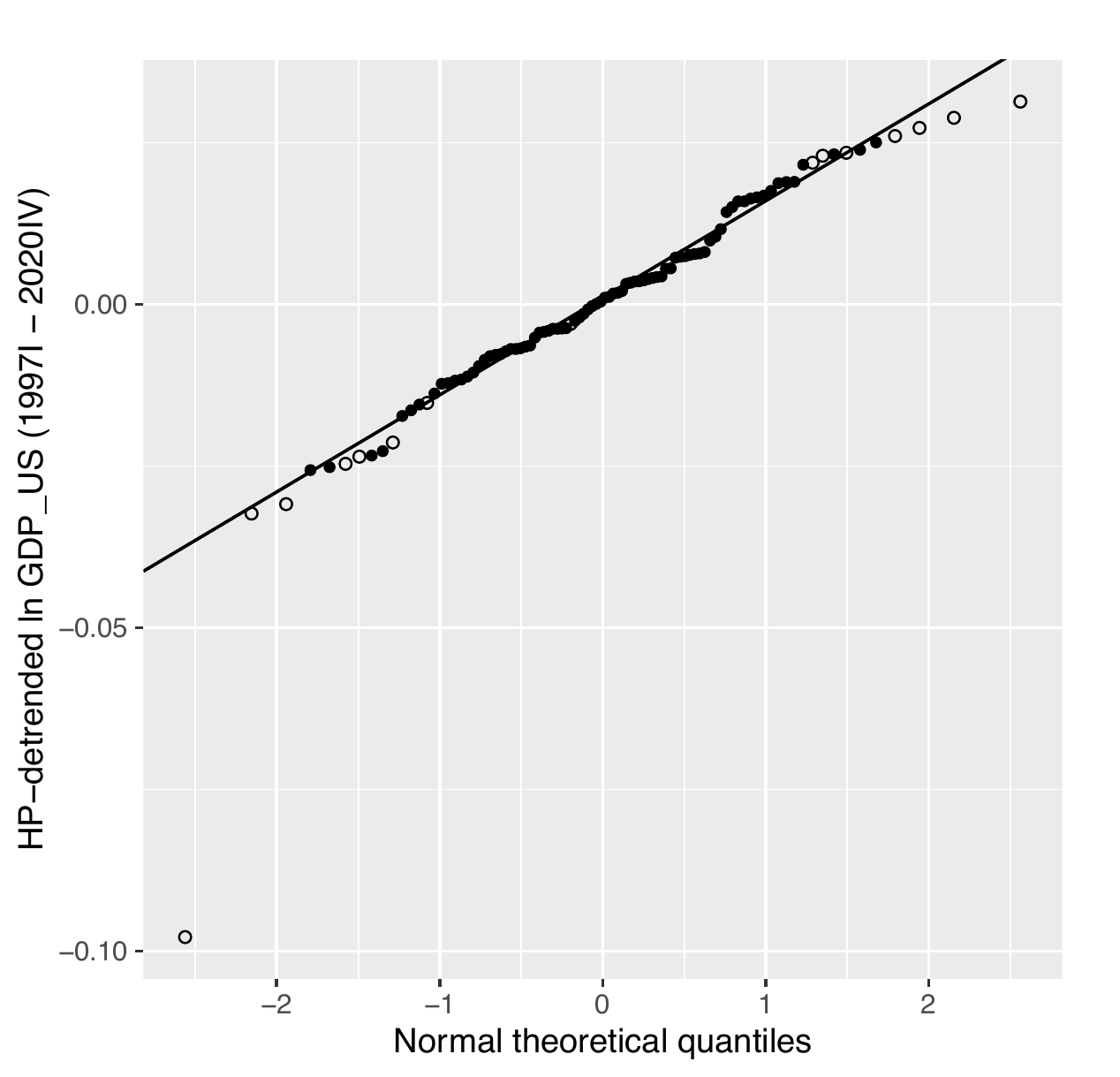}
\includegraphics[width=0.495\textwidth]{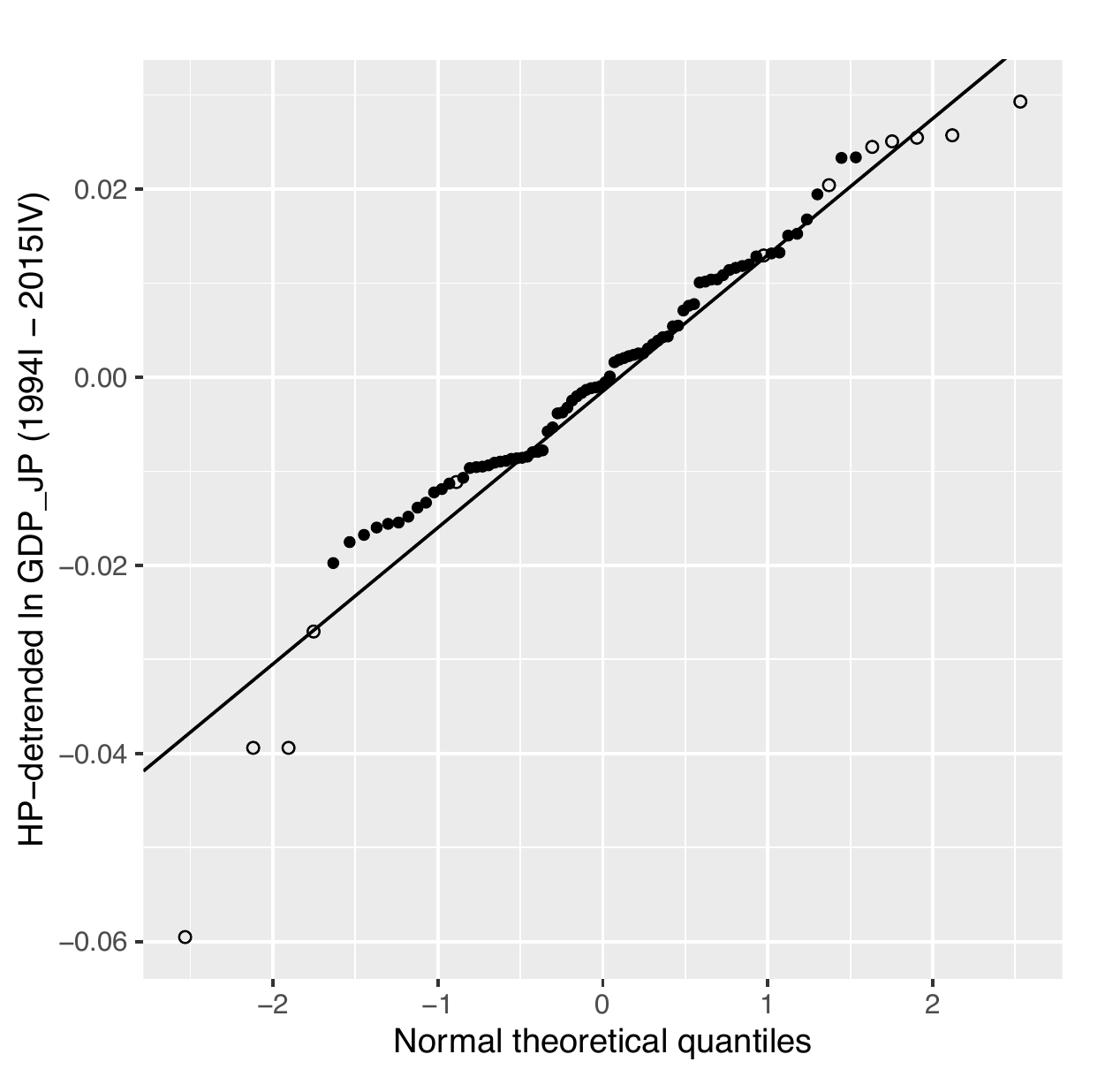}
\caption{
Quantile-quantile plots of recent (US: 1997I--2020IV; Japan: 1994I--2015IV) quarterly HP-detrended log GDP against the normal distribution for the US (left) and Japan (right). 
The time periods correspond to those when the sectoral elasticities are estimated.
Open dots indicate 2007I--2009IV (GFC) and 2019IV-- (COVID-19 pandemic) observations. 
Source: \citet{fred, co}.
} \label{figx}
\end{figure}

In the CES economy with a sector-specific elasticity of substitution,
we use the Domar aggregator of general type (\ref{domarces}) with
recursion.   
The right panel of Figure \ref{fig2} depicts the resulting QQ plot of the aggregate fluctuations generated by zero-mean normal shocks.
In this case, we observe positive expected output growth i.e., $\overline{\ln H} =1.10\%$.
The value demonstrates the robustness of the elastic CES economy
relative to the Cobb-Douglas economy.
We also observe that normal shocks to the elastic CES economy result in aggregate fluctuations with tail asymmetry similar to that depicted in the right panel of Figure \ref{fig0}.
For sake of credibility, we show in Figure \ref{figx} (right) the empirical aggregate output fluctuations focusing on the period 1994--2015 from which our sectoral elasticities are estimated.
It is obvious that extreme observations belong to periods around the GFC (global financial crisis), which was a massive external (non-sectoral) shock to the Japanese economy.
The plot otherwise appears rather positively skewed, as predicted by our empirical result ($\bar{\hat{\sigma}}=1.54$).

Figure \ref{figx} (left) shows the aggregate output fluctuations focusing on the period 1997--2020 from which we estimated sectoral elasticities for the US (see \ref{app2} for details).
In this case, the extreme observations also belong to periods around the GFC and COVID-19 pandemic.
The plot, however, seems to be in a straight line, which is consistent with our empirical result ($\bar{\hat{\sigma}}=1.08$). 
Our elasticity estimates for the US economy also coincide with the
estimates of the elasticity of substitution across intermediate inputs ($\epsilon_M\text{ all}= 1.05$) by \citet{MPY} based on 1997--2007 input-output accounts.
These elasticity estimates, however, differ from those obtained by \citet{atalay} based on 1997--2013 input-output accounts that \citet{baqaee} employed in their simulation ($\varepsilon = 0.001$).
An inelastic economy as such is rather consistent with negatively skewed aggregate fluctuations spanning the postwar US economy as depicted in Figure \ref{fig0} (left) than those of recent times depicted in Figure \ref{figx} (left).

\section{Concluding Remarks}

\citet{aceAER}'s claim was that a heavy tailed aggregate fluctuation can emerge from heavy tailed microeconomic shocks because of the network heterogeneity of the input-output linkages (which will be fixed under Cobb-Douglas economy), even if the central limit theorem implies that the aggregate fluctuation must converge into a normal distribution. 
Our findings that the US economy in recent years has been essentially a Cobb-Douglas one, and that recent aggregate fluctuations exhibit a tail risk as captured in Figure \ref{figx} (left), therefore, indicate a negatively skewed distribution of microeconomic shocks. 
In the meanwhile, our finding of Japan's elastic economy, together with the assumption of negatively skewed microeconomic shocks, provide a better understanding of the peculiar pattern of its recent aggregate fluctuations as captured in Figure \ref{figx} (right).

It is well documented that the Japanese have been more creative in discovering how to produce than in what to produce.
The empirical results obtained in this study provide some evidence to believe that such a spirit is engraved in the nation's economy.
Undeniably, the technologies embodied in a production function have been acquired over the long course of research and development.
Japan must have developed its elastic economy through the grinding process of discovering more efficient and inexpensive ways to produce while overcoming the many external turbulences it confronted.
Whatever the cause may be, an elastic economy equipped with many substitutable technologies must be favourable with respect to robustness against turbulence.
Ultimately, human creativity expands the production function in two dimensions: productivity and substitutability, and the elasticity of substitution in particular, 
which brings synergism between the economic entities, must be worthy of further investigation.

\def\thesection{Appendix 1}
\section{GBM Property of Sectoral Productivity \label{app1}}
A geometric Brownian motion (GBM) can be specified by the following stochastic differential equation (SDE):
\begin{align}
\mathrm{d} X_t = \mu X_t \mathrm{d} t + \sigma X_t\mathrm{d}B_t 
\end{align}
where, $\mu$ denotes the drift parameter, $\sigma$ denotes the volatility, and $B_t \sim \mathcal{N}(0, t)$.
Ito's Lemma implies that the above SDE is equivalent to the following:
\begin{align}
\mathrm{d} \ln X_t = (\mu - \sigma^2/2) \mathrm{d}t + \sigma \mathrm{d}B_t
\end{align}
where this SDE is solvable by integration. 
The solution follows below:
\begin{align}
\ln {X_T} = \ln X_0 + (\mu - \sigma^2/2) T + \sigma B_T
\end{align}
Since $B_T \sim \mathcal{N}(0, T)$, the first and the second moments for $\ln  ({X_T}/{X_0})$ can be evaluated as follows:
\begin{align}
\mathbb{E} \left[ \ln ({X_T}/{X_0}) \right] = (\mu - \sigma^2/2) T,
&&
\mathrm{var} \left[ \ln ({X_T}/{X_0}) \right] = \sigma^2 T
\label{ev}
\end{align}

There are several ways of estimating the volatility and the drift parameters of a GBM empirically from $\ell$ size of historical data $(X_1, \cdots, X_{\ell})$.
The obvious approach is the following, which is based on the sample moments:
\begin{align}
\hat{\sigma} = \sqrt{\frac{1}{\ell -2}\sum_{k=1}^{\ell-1} \left(\Delta \ln X_k - \frac{\sum_{k=1}^{\ell -1} \Delta \ln X_k}{\ell -1} \right)^2},
&&
~~~~\hat{\mu} = \frac{\sum_{k=1}^{\ell-1} \Delta \ln X_k}{\ell-1} + \frac{1}{2} \hat{\sigma}^2
\end{align}
Alternatively, \citet{hlm} devised parameter estimates of the following based on the simulated maximum likelihood method.
\begin{align}
\hat{\sigma} = \sqrt{\frac{1}{\ell -1}\sum_{k=1}^{\ell-1} \left(\frac{X_{k+1}}{X_k} - \frac{\sum_{k=1}^{\ell -1} \frac{X_{k+1}}{X_{k}}}{\ell -1}  \right)^2},
&&
~~~~\hat{\mu} = \frac{\sum_{k=1}^{\ell -1} \frac{X_{k+1}}{X_{k}}}{\ell -1} -1
\label{dlm}
\end{align}
The two methods produce very similar results for our data.
Table \ref{tab2} summarizes the estimated annual volatilities and drift parameters for all 100 sectors by formula (\ref{dlm}) using the historical data of sectoral annual TFP from 1995 to 2015.
Moreover, conforming to \citet{engeco}, we check the normality of the annual TFP growth rates using the Shapiro-Wilk W test.
Normality was rejected in 19 out of the 100 sectors.
The annual volatility, with a t-statistic over 2, ranges from 0.260 to 0.523, whereas the simple average concerning all 100 sectors is 0.251.

\begin{center}
\begin{ThreePartTable}
\begin{TableNotes}
\footnotesize
\item[Note:] 
The simple mean of all annual volatilities is 0.251.
The normality of TFP growth rates is examined by the Shapiro-Wilk W test, where rejection of normality is indicated by the label `no', and a blank is left otherwise. 
\item[*1] Electronic data processing machines, digital and analog computer equipment and accessories
\item[*2] Image information, sound information and character information production
\end{TableNotes}
\newcolumntype{.}{D{.}{.}{3}}
\begin{longtable}{c@{~~} l@{~~} .@{~~} .@{~} c}
\caption{GBM property of estimated productivity growth for all sectors.} \label{tab2} \\
\hline\noalign{\smallskip}	
\multicolumn{1}{c}{id} & 
\multicolumn{1}{c}{sector} &
\multicolumn{1}{c}{$\hat{\mu}$} &
\multicolumn{1}{c}{$\hat{\sigma}$} &
\multicolumn{1}{c}{norm.} \\
\hline\noalign{\smallskip}
\endfirsthead
\multicolumn{5}{c}%
{{\tablename\ \thetable{} -- continued from previous page}} \\
\hline\noalign{\smallskip}	
\multicolumn{1}{c}{id} & 
\multicolumn{1}{c}{sector} &
\multicolumn{1}{c}{$\hat{\mu}$} &
\multicolumn{1}{c}{$\hat{\sigma}$} &
\multicolumn{1}{c}{norm.} \\ \hline\noalign{\smallskip}
\endhead
\hline
\endfoot
\hline 
\insertTableNotes
\endlastfoot
1	&	Agriculture	&	0.015	&	0.051	&		\\
2	&	Agricultural services	&	0.000	&	0.036	&		\\
3	&	Forestry	&	0.021	&	0.053	&		\\
4	&	Fisheries	&	0.001	&	0.065	&		\\
5	&	Mining	&	0.013	&	0.042	&		\\
6	&	Livestock products	&	0.010	&	0.118	&		\\
7	&	Seafood products	&	$-$0.002	&	0.060	&		\\
8	&	Flour and grain mill products	&	0.128	&	0.777	&		\\
9	&	Miscellaneous foods and related products	&	$-$0.016	&	0.073	&		\\
10	&	Beverages	&	$-$0.003	&	0.058	&		\\
11	&	Prepared animal foods and organic fertilizers	&	$-$0.023	&	0.113	&		\\
12	&	Tobacco	&	$-$0.072	&	0.222	&		\\
13	&	Textile products (except chemical fibers)	&	0.019	&	0.242	&		\\
14	&	Chemical fibers	&	$-$0.060	&	0.279	&		\\
15	&	Pulp, paper, and coated and glazed paper	&	$-$0.012	&	0.143	&		\\
16	&	Paper products	&	$-$0.071	&	0.576	&	no	\\
17	&	Chemical fertilizers	&	$-$0.016	&	0.040	&		\\
18	&	Basic inorganic chemicals	&	$-$0.021	&	0.097	&	no	\\
19	&	Basic organic chemicals	&	$-$0.043	&	0.146	&	no	\\
20	&	Organic chemicals	&	$-$0.027	&	0.078	&		\\
21	&	Pharmaceutical products	&	0.192	&	1.362	&		\\
22	&	Miscellaneous chemical products	&	0.046	&	0.209	&		\\
23	&	Petroleum products	&	$-$0.057	&	0.300	&	no	\\
24	&	Coal products	&	$-$0.063	&	0.259	&	no	\\
25	&	Glass and its products	&	0.027	&	0.341	&	no	\\
26	&	Cement and its products	&	$-$0.003	&	0.054	&		\\
27	&	Pottery	&	0.015	&	0.059	&		\\
28	&	Miscellaneous ceramic, stone and clay products	&	0.001	&	0.067	&		\\
29	&	Pig iron and crude steel	&	$-$0.021	&	0.074	&		\\
30	&	Miscellaneous iron and steel	&	$-$0.049	&	0.292	&		\\
31	&	Smelting and refining of non$-$ferrous metals	&	$-$0.042	&	0.160	&		\\
32	&	Non$-$ferrous metal products	&	$-$0.113	&	0.272	&		\\
33	&	Fabricated constructional and architectural metal products	&	$-$0.053	&	0.220	&		\\
34	&	Miscellaneous fabricated metal products	&	0.405	&	1.638	&		\\
35	&	General$-$purpose machinery	&	$-$0.066	&	0.196	&		\\
36	&	Production machinery	&	$-$0.056	&	0.350	&		\\
37	&	Office and service industry machines	&	0.112	&	0.221	&	no	\\
38	&	Miscellaneous business oriented machinery	&	0.141	&	0.379	&		\\
39	&	Ordnance	&	0.024	&	0.107	&		\\
40	&	Semiconductor devices and integrated circuits	&	0.072	&	0.276	&	no	\\
41	&	Miscellaneous electronic components and devices	&	0.310	&	1.641	&	no	\\
42	&	Electrical devices and parts	&	0.032	&	1.509	&		\\
43	&	Household electric appliances	&	0.018	&	0.134	&		\\
44	&	Electronic equipment and electric measuring instruments	&	0.051	&	0.147	&		\\
45	&	Miscellaneous electrical machinery equipment	&	0.008	&	0.070	&		\\
46	&	Image and audio equipment	&	$-$0.136	&	0.523	&	no	\\
47	&	Communication equipment	&	0.138	&	1.706	&		\\
48	&	Electronic data processing machines, etc*1	&	0.127	&	0.980	&		\\
49	&	Motor vehicles (including motor vehicles bodies)	&	$-$0.028	&	0.305	&		\\
50	&	Motor vehicle parts and accessories	&	0.060	&	0.314	&		\\
51	&	Other transportation equipment	&	$-$0.059	&	0.584	&		\\
52	&	Printing	&	$-$0.005	&	0.078	&		\\
53	&	Lumber and wood products	&	0.015	&	0.073	&		\\
54	&	Furniture and fixtures	&	0.008	&	0.040	&	no	\\
55	&	Plastic products	&	$-$0.011	&	0.097	&		\\
56	&	Rubber products	&	$-$0.053	&	0.206	&		\\
57	&	Leather and leather products	&	0.009	&	0.057	&	no	\\
58	&	Watches and clocks	&	0.014	&	0.088	&		\\
59	&	Miscellaneous manufacturing industries	&	$-$0.003	&	0.091	&		\\
60	&	Electricity	&	0.000	&	0.053	&		\\
61	&	Gas, heat supply	&	$-$0.016	&	0.070	&		\\
62	&	Waterworks	&	$-$0.090	&	0.333	&		\\
63	&	Water supply for industrial use	&	$-$0.010	&	0.036	&		\\
64	&	Sewage disposal	&	$-$0.002	&	0.028	&		\\
65	&	Waste disposal	&	0.014	&	0.059	&		\\
66	&	Construction	&	$-$0.040	&	0.078	&		\\
67	&	Civil engineering	&	0.007	&	0.034	&		\\
68	&	Wholesale	&	0.008	&	0.099	&		\\
69	&	Retail	&	0.023	&	0.035	&		\\
70	&	Railway	&	0.107	&	0.488	&		\\
71	&	Road transportation	&	0.133	&	1.430	&		\\
72	&	Water transportation	&	$-$0.010	&	0.044	&		\\
73	&	Air transportation	&	$-$0.014	&	0.053	&	no	\\
74	&	Other transportation and packing	&	0.016	&	0.071	&		\\
75	&	Mail	&	0.036	&	0.229	&	no	\\
76	&	Hotels	&	0.018	&	0.061	&		\\
77	&	Eating and drinking services	&	0.014	&	0.076	&		\\
78	&	Communications	&	0.077	&	0.173	&		\\
79	&	Broadcasting	&	0.024	&	0.531	&		\\
80	&	Information services	&	$-$0.006	&	0.132	&		\\
81	&	Image information, etc*2	&	$-$0.050	&	0.102	&		\\
82	&	Finance	&	0.032	&	0.193	&		\\
83	&	Insurance	&	0.014	&	0.093	&		\\
84	&	Housing	&	$-$0.108	&	0.284	&	no	\\
85	&	Real estate	&	0.038	&	0.049	&		\\
86	&	Research	&	$-$0.013	&	0.076	&		\\
87	&	Advertising	&	0.015	&	0.047	&		\\
88	&	Rental of office equipment and goods	&	0.028	&	0.197	&		\\
89	&	Automobile maintenance services	&	0.000	&	0.066	&		\\
90	&	Other services for businesses	&	$-$0.006	&	0.058	&	no	\\
91	&	Public administration	&	0.001	&	0.069	&		\\
92	&	Education	&	0.016	&	0.026	&		\\
93	&	Medical service, health and hygiene	&	$-$0.001	&	0.072	&		\\
94	&	Social insurance and social welfare	&	0.031	&	0.061	&		\\
95	&	Nursing care	&	0.002	&	0.098	&	no	\\
96	&	Entertainment	&	0.011	&	0.071	&		\\
97	&	Laundry, beauty and bath services	&	0.013	&	0.143	&	no	\\
98	&	Other services for individuals	&	0.050	&	0.123	&		\\
99	&	Membership organizations	&	0.030	&	0.118	&		\\
100	&	Activities not elsewhere classified	&	0.101	&	0.288	&	no	\\

\end{longtable}
\end{ThreePartTable}
\end{center}

\def\thesection{Appendix 2}
\section{Sectoral elasticities of substitution for the US economy \label{app2}}
This section is devoted to our estimation of sectoral substitution elasticities for the US, in the same manner as we did for Japan.
First, we create $n\times n$ input-output tables using the make and use tables of $n = 71$ industries in nominal terms for 24 years (1997--2020), available at \citet{bea}.
Next, we create tables in real terms by using price indices available as chain-type price indexes for gross output by industry.
Note that the real value added of an industry is estimated by double deflation, so that price indices for value added can be derived from nominal and real value added accounts.
As for instruments, we utilize the integrated multifactor productivity
(MFP), taken from the 1987--2019 Production Account Capital Table
\citep{bls} of the BEA-BLS Integrated Industry-level Production
Accounts (KLEMS), for $n$ factor inputs.\footnote{We use the last
  three years' (2017, 2018, 2019) average of the accounts to
  instrument for the 2020 explanatory variable.}
To instrument primary factor prices, we apply three different
instruments, namely, total factor productivity (i.e., aggregate TFP),
a capital price deflator, and a labor price deflator, obtainable from the Annual total factor productivity and related measure for major sectors \citep{bls}.
Thus, our instrumental variables are $v^a$ (sectoral MFP with aggregate TFP), $v^b$ (sectoral MFP with capital price deflator), and $v^c$ (sectoral MFP with labor price deflator), all of which are of $n+1$ dimension.

Estimation of sectoral elasticities of substitution was conducted according to the estimation framework presented in section 3.
The results are summarized in Table \ref{tab:long3}.
The first column (LS FE) reports the least squares fixed effects estimation results, without instrumenting for the explanatory variable.
The second column (IV FE) reports the instrumental variable fixed effects estimation results, using the IVs reported in the last column.
In all cases, overidentification tests are not rejected.
Moreover, first-stage F values are large enough.
The estimates for the elasticity of substitution $\hat{\sigma}$ are larger when IVs are applied.
According to the endogeneity test results, we accept the LS FE
estimates for sector ids 6, 9, 10, 11, 21, 34, 37, 41, 48, 55, 56, 62,
63, 64, 65, and 71, instead of the IV FE estimates.
The simple mean of the estimated (accepted) elasticities is $\bar{\hat{\sigma}}=1.08$.

\begin{center}
\begin{ThreePartTable}
\begin{TableNotes}
\footnotesize
\item[Notes:] The id number corresponds to the numerical position of an industry of the input-output table of 71 industries \citet{bea}.
Values in parentheses indicate p-values.
\item[*1] First-stage (Cragg-Donald Wald) F statistic for 2SLS FE estimation.
The rule of thumb to reject the hypothesis that the explanatory variable is only weakly correlated with the instrument is for this to exceed 10.
\item[*2] Overidentification test by Sargan statistic.
Rejection of the null indicates that the instruments are correlated with the residuals.
\item[*3] Endogeneity test by Davidson-MacKinnon F statistic.
Rejection of the null indicates that the instrumental variables fixed effects estimator should be employed.
\item[*4] Instrumental variables applied, where 1, 2, and 3, indicate
  ${v^a}$, ${v^b}$, and ${v^c}$, respectively, and l, f, and d
  indicate first lag, first forward, and first difference, respectively.
\end{TableNotes}
\newcolumntype{.}{D{.}{.}{3}}
\newcolumntype{i}{D{.}{}{0}}
\begin{longtable}{r@{~~~}.@{~~~} . .@{~~~} .@{~} i .@{} .@{~~~~} .@{} . c}
\caption{Estimation of the elasticity of substitution for 71 US sectors.} \label{tab:long3} \\
\hline\noalign{\smallskip}	
& \multicolumn{2}{c}{LS FE} & \multicolumn{8}{c}{IV FE} \\
\cmidrule(r){2-3}\cmidrule(r){4-11}
\multicolumn{1}{c}{id} &\multicolumn{1}{c}{$\hat{\sigma}$ } &\multicolumn{1}{c}{s.e.} &\multicolumn{1}{c}{$\hat{\sigma}$} &\multicolumn{1}{c}{s.e.} &\multicolumn{1}{c}{1st F\tnote{*1}} &\multicolumn{2}{c}{Overid.\tnote{*2}} &\multicolumn{2}{c}{Endog.\tnote{*3}}  &\multicolumn{1}{c}{IVs\tnote{*4}} \\ \hline\noalign{\smallskip}
\endfirsthead
\multicolumn{11}{c}%
{{\tablename\ \thetable{} -- continued from previous page}} \\
\hline\noalign{\smallskip}	
& \multicolumn{2}{c}{LS FE} & \multicolumn{8}{c}{IV FE} \\
\cmidrule(r){2-3}\cmidrule(r){4-11}
\multicolumn{1}{c}{id} &\multicolumn{1}{c}{$\hat{\sigma}$ } &\multicolumn{1}{c}{s.e.} &\multicolumn{1}{c}{$\hat{\sigma}$} &\multicolumn{1}{c}{s.e.} &\multicolumn{1}{c}{1st F\tnote{*1}} &\multicolumn{2}{c}{Overid.\tnote{*2}} &\multicolumn{2}{c}{Endog.\tnote{*3}}  &\multicolumn{1}{c}{IVs\tnote{*4}} \\ \hline\noalign{\smallskip}
\endhead
\hline
\endfoot
\hline 
\insertTableNotes
\endlastfoot
1	&	0.604		&	0.082		&	1.261		&	0.152		&	309		&	1.48		&	(0.223)		&	27.75		&	(0.000)		&	1, 2		\\
2	&	0.532		&	0.101		&	1.348		&	0.187		&	302		&	0.10		&	(0.755)		&	32.42		&	(0.000)		&	l1, 2		\\
3	&	0.533		&	0.087		&	1.124		&	0.182		&	197		&	1.14		&	(0.286)		&	15.67		&	(0.000)		&	1, l2		\\
4	&	0.635		&	0.083		&	0.957		&	0.138		&	388		&	0.25		&	(0.616)		&	8.56		&	(0.003)		&	2, 1		\\
5	&	1.143		&	0.085		&	1.549		&	0.161		&	298		&	0.03		&	(0.868)		&	11.76		&	(0.001)		&	l2, 1		\\
6	&	0.812		&	0.106		&	0.749		&	0.190		&	313		&	0.03		&	(0.859)		&	0.16		&	(0.689)		&	1, 2		\\
7	&	0.611		&	0.085		&	1.153		&	0.167		&	261		&	0.53		&	(0.469)		&	16.25		&	(0.000)		&	l1, 2		\\
8	&	0.748		&	0.094		&	1.495		&	0.175		&	306		&	0.32		&	(0.571)		&	29.84		&	(0.000)		&	l1, 2		\\
9	&	0.843		&	0.081		&	0.861		&	0.146		&	314		&	0.21		&	(0.649)		&	0.02		&	(0.878)		&	1, 2		\\
10	&	1.037		&	0.079		&	0.746		&	0.151		&	262		&	0.68		&	(0.411)		&	3.82		&	(0.051)		&	1, l2		\\
11	&	0.888		&	0.077		&	1.037		&	0.079		&	291		&	1.71		&	(0.190)		&	2.06		&	(0.151)		&	l1, 2		\\
12	&	0.831		&	0.082		&	1.016		&	0.141		&	300		&	0.97		&	(0.324)		&	5.09		&	(0.024)		&	l1, 2		\\
13	&	1.162		&	0.074		&	1.063		&	0.150		&	147		&	0.95		&	(0.330)		&	7.91		&	(0.005)		&	l1, 2		\\
14	&	0.996		&	0.075		&	0.728		&	0.174		&	298		&	1.06		&	(0.304)		&	17.02		&	(0.000)		&	l1, 2		\\
15	&	0.801		&	0.067		&	1.434		&	0.136		&	319		&	1.86		&	(0.173)		&	19.51		&	(0.000)		&	l1, 2		\\
16	&	0.901		&	0.091		&	1.211		&	0.120		&	290		&	0.35		&	(0.554)		&	8.07		&	(0.005)		&	l1, 2		\\
17	&	0.834		&	0.085		&	1.231		&	0.167		&	306		&	0.89		&	(0.346)		&	10.19		&	(0.001)		&	l1, 2		\\
18	&	0.906		&	0.070		&	1.203		&	0.156		&	319		&	1.57		&	(0.211)		&	21.21		&	(0.000)		&	l1, 2		\\
19	&	0.844		&	0.072		&	1.353		&	0.127		&	279		&	2.19		&	(0.139)		&	20.81		&	(0.000)		&	l1, 2		\\
20	&	0.695		&	0.090		&	1.340		&	0.141		&	313		&	1.29		&	(0.256)		&	12.73		&	(0.000)		&	l1, 2		\\
21	&	0.506		&	0.123		&	1.126		&	0.166		&	351		&	0.67		&	(0.413)		&	0.55		&	(0.458)		&	l1, 2		\\
22	&	0.730		&	0.087		&	0.536		&	0.212		&	297		&	2.14		&	(0.144)		&	4.23		&	(0.040)		&	l1, 2		\\
23	&	0.677		&	0.078		&	0.944		&	0.166		&	366		&	2.13		&	(0.145)		&	5.35		&	(0.021)		&	1, 2		\\
24	&	0.768		&	0.095		&	0.930		&	0.135		&	167		&	2.01		&	(0.156)		&	5.44		&	(0.020)		&	l1, 2		\\
25	&	0.530		&	0.088		&	0.288		&	0.226		&	268		&	0.18		&	(0.674)		&	12.00		&	(0.001)		&	l1, 2		\\
26	&	0.703		&	0.087		&	0.954		&	0.174		&	302		&	1.45		&	(0.229)		&	19.79		&	(0.000)		&	l1, 2		\\
27	&	0.740		&	0.062		&	1.253		&	0.165		&	307		&	1.77		&	(0.183)		&	34.09		&	(0.000)		&	l1, 2		\\
28	&	0.833		&	0.081		&	1.237		&	0.112		&	324		&	0.02		&	(0.887)		&	8.88		&	(0.003)		&	l1, 2		\\
29	&	0.764		&	0.066		&	1.147		&	0.144		&	279		&	0.04		&	(0.849)		&	8.70		&	(0.003)		&	l1, 2		\\
30	&	0.956		&	0.102		&	1.028		&	0.125		&	315		&	0.07		&	(0.796)		&	13.84		&	(0.000)		&	l1, 2		\\
31	&	0.721		&	0.064		&	1.461		&	0.186		&	320		&	1.16		&	(0.281)		&	17.47		&	(0.000)		&	l1, 2		\\
32	&	0.279		&	0.146		&	1.067		&	0.114		&	276		&	0.16		&	(0.688)		&	18.20		&	(0.000)		&	l1, 2		\\
33	&	0.720		&	0.074		&	1.214		&	0.271		&	279		&	1.97		&	(0.161)		&	24.69		&	(0.000)		&	l1, d2		\\
34	&	0.765		&	0.093		&	1.255		&	0.140		&	91		&	0.51		&	(0.475)		&	2.98		&	(0.084)		&	d1, d2		\\
35	&	0.584		&	0.114		&	1.168		&	0.257		&	289		&	0.16		&	(0.689)		&	24.71		&	(0.000)		&	l1, 2		\\
36	&	0.860		&	0.116		&	1.416		&	0.215		&	273		&	0.78		&	(0.377)		&	10.90		&	(0.001)		&	l1, 2		\\
37	&	0.083		&	0.139		&	1.377		&	0.216		&	256		&	1.24		&	(0.266)		&	1.70		&	(0.192)		&	l1, 2		\\
38	&	0.621		&	0.104		&	0.217		&	0.258		&	262		&	0.07		&	(0.797)		&	20.22		&	(0.000)		&	l2, 3		\\
39	&	0.608		&	0.106		&	1.348		&	0.208		&	318		&	0.57		&	(0.450)		&	7.61		&	(0.006)		&	l1, 2		\\
40	&	0.911		&	0.096		&	0.996		&	0.188		&	304		&	0.00		&	(0.994)		&	44.01		&	(0.000)		&	l1, 1		\\
41	&	0.993		&	0.089		&	1.862		&	0.183		&	308		&	0.18		&	(0.671)		&	2.48		&	(0.115)		&	1, l2		\\
42	&	0.913		&	0.073		&	0.735		&	0.167		&	340		&	1.85		&	(0.174)		&	58.21		&	(0.000)		&	2, l3		\\
43	&	0.773		&	0.094		&	1.675		&	0.132		&	330		&	0.09		&	(0.768)		&	5.19		&	(0.023)		&	l1, 2		\\
44	&	0.762		&	0.084		&	1.072		&	0.171		&	293		&	0.11		&	(0.741)		&	38.11		&	(0.000)		&	l1, d2		\\
45	&	1.120		&	0.101		&	1.514		&	0.153		&	219		&	1.58		&	(0.209)		&	31.97		&	(0.000)		&	l2, l3		\\
46	&	0.804		&	0.126		&	2.055		&	0.203		&	333		&	1.49		&	(0.223)		&	4.13		&	(0.042)		&	l1, 2		\\
47	&	0.691		&	0.116		&	1.154		&	0.223		&	235		&	1.32		&	(0.251)		&	28.22		&	(0.000)		&	l1, 2		\\
48	&	1.226		&	0.269		&	1.664		&	0.233		&	326		&	0.00		&	(0.993)		&	0.06		&	(0.808)		&	1, l2		\\
49	&	0.627		&	0.093		&	1.063		&	0.481		&	326		&	0.72		&	(0.397)		&	48.50		&	(0.000)		&	l1, 2		\\
50	&	0.743		&	0.075		&	1.507		&	0.170		&	304		&	1.97		&	(0.160)		&	28.24		&	(0.000)		&	l1, 2		\\
51	&	0.738		&	0.105		&	1.290		&	0.139		&	288		&	0.26		&	(0.613)		&	8.40		&	(0.004)		&	l1, 1		\\
52	&	0.919		&	0.077		&	1.186		&	0.192		&	350		&	2.30		&	(0.130)		&	72.85		&	(0.000)		&	l1, 3		\\
53	&	0.524		&	0.068		&	1.758		&	0.135		&	323		&	0.65		&	(0.419)		&	14.11		&	(0.000)		&	l1, 2		\\
54	&	0.718		&	0.087		&	0.855		&	0.126		&	304		&	0.62		&	(0.431)		&	4.31		&	(0.038)		&	l1, 1		\\
55	&	0.787		&	0.081		&	0.929		&	0.152		&	277		&	0.82		&	(0.365)		&	1.96		&	(0.161)		&	l1, l2		\\
56	&	1.236		&	0.083		&	0.902		&	0.153		&	284		&	2.40		&	(0.121)		&	1.83		&	(0.176)		&	l1, 2		\\
57	&	0.566		&	0.076		&	1.245		&	0.148		&	307		&	1.82		&	(0.178)		&	28.10		&	(0.000)		&	l1, 1		\\
58	&	1.075		&	0.079		&	1.154		&	0.142		&	333		&	1.27		&	(0.260)		&	44.26		&	(0.000)		&	l1, 1		\\
59	&	0.814		&	0.093		&	1.814		&	0.145		&	314		&	0.04		&	(0.849)		&	46.53		&	(0.000)		&	l1, 2		\\
60	&	0.765		&	0.094		&	1.753		&	0.179		&	308		&	1.99		&	(0.159)		&	11.45		&	(0.001)		&	1, l2		\\
61	&	0.683		&	0.084		&	1.235		&	0.177		&	298		&	0.52		&	(0.470)		&	17.88		&	(0.000)		&	1, l2		\\
62	&	0.270		&	0.091		&	1.167		&	0.155		&	309		&	0.10		&	(0.758)		&	0.76		&	(0.383)		&	1, l2		\\
63	&	0.667		&	0.076		&	0.364		&	0.170		&	325		&	0.83		&	(0.361)		&	2.11		&	(0.147)		&	1, l2		\\
64	&	0.536		&	0.085		&	0.795		&	0.138		&	296		&	0.00		&	(0.965)		&	0.00		&	(0.965)		&	l1, 2		\\
65	&	0.522		&	0.061		&	0.494		&	0.158		&	291		&	0.60		&	(0.438)		&	2.21		&	(0.138)		&	l2, 1		\\
66	&	0.549		&	0.073		&	0.604		&	0.113		&	320		&	0.78		&	(0.378)		&	72.74		&	(0.000)		&	l1, 1		\\
67	&	0.871		&	0.084		&	1.439		&	0.139		&	309		&	3.26		&	(0.071)		&	20.85		&	(0.000)		&	l1, 1		\\
68	&	0.743		&	0.103		&	1.371		&	0.150		&	297		&	1.15		&	(0.284)		&	4.38		&	(0.036)		&	f1, 2		\\
69	&	0.209		&	0.126		&	1.083		&	0.192		&	293		&	1.25		&	(0.264)		&	4.73		&	(0.030)		&	l1, 3		\\
70	&	0.606		&	0.068		&	0.583		&	0.221		&	324		&	0.96		&	(0.327)		&	25.66		&	(0.000)		&	l1, 1		\\
71	&	0.660		&	0.086		&	1.088		&	0.130		&	303		&	0.10		&	(0.752)		&	1.44		&	(0.230)		&	l1, 1		\\

\end{longtable}
\end{ThreePartTable}
\end{center}

\section*{Acknowledgements}
The authors would like to thank the editor and the anonymous reviewers for their helpful comments and suggestions. \\
JSPS Kakenhi Grant numbers: 19H04380, 20K22139 \\
The authors declare that they have no conflict of interest.

\bibliographystyle{spbasic_x}      
{\raggedright
\bibliography{bibfile}   

\begin{thebibliography}{28}
\providecommand{\natexlab}[1]{#1}
\providecommand{\url}[1]{{#1}}
\providecommand{\urlprefix}{URL }
\expandafter\ifx\csname urlstyle\endcsname\relax
  \providecommand{\doi}[1]{DOI~\discretionary{}{}{}#1}\else
  \providecommand{\doi}{DOI~\discretionary{}{}{}\begingroup
  \urlstyle{rm}\Url}\fi
\providecommand{\eprint}[2][]{\url{#2}}

\bibitem[{Acemoglu and Azar(2020)}]{ace2020}
Acemoglu D, Azar PD (2020) Endogenous production networks. Econometrica
  88(1):33--82,
  \href{http://dx.doi.org/10.3982/ECTA15899}{\doi{10.3982/ECTA15899}}

\bibitem[{Acemoglu et~al(2012)Acemoglu, Carvalho, Ozdaglar, and
  Tahbaz-Salehi}]{aceECTA}
Acemoglu D, Carvalho VM, Ozdaglar A, Tahbaz-Salehi A (2012) The network origins
  of aggregate fluctuations. Econometrica 80(5):1977--2016,
  \href{http://dx.doi.org/10.3982/ECTA9623}{\doi{10.3982/ECTA9623}}

\bibitem[{Acemoglu et~al(2017)Acemoglu, Ozdaglar, and Tahbaz-Salehi}]{aceAER}
Acemoglu D, Ozdaglar A, Tahbaz-Salehi A (2017) Microeconomic origins of
  macroeconomic tail risks. American Economic Review 107(1):54--108,
  \href{http://dx.doi.org/10.1257/aer.20151086}{\doi{10.1257/aer.20151086}}

\bibitem[{Atalay(2017)}]{atalay}
Atalay E (2017) How important are sectoral shocks? American Economic Journal:
  Macroeconomics 9(4):254--80,
  \href{http://dx.doi.org/10.1257/mac.20160353}{\doi{10.1257/mac.20160353}}

\bibitem[{Baqaee and Farhi(2019)}]{baqaee}
Baqaee DR, Farhi E (2019) The macroeconomic impact of microeconomic shocks:
  Beyond hulten's theorem. Econometrica 87(4):1155--1203,
  \href{http://dx.doi.org/10.3982/ECTA15202}{\doi{10.3982/ECTA15202}}

\bibitem[{BEA(2022)}]{bea}
BEA (2022) {Bureau of Economic Analysis, Input-Output Accounts Data.}
  \urlprefix\url{https://www.bea.gov/industry/input-output-accounts-data}

\bibitem[{BLS(2022)}]{bls}
BLS (2022) {U.S. Bureau of Labor Statistics, Office of Productivity and
  Technology}. \urlprefix\url{https://www.bls.gov/productivity/tables/}

\bibitem[{{Cabinet Office}(2021)}]{co}
{Cabinet Office} (2021) Quarterly estimates of gdp - release archive.
  \urlprefix\url{https://www.esri.cao.go.jp/en/sna/data/sokuhou/files/toukei_top.html}

\bibitem[{Carvalho et~al(2020)Carvalho, Nirei, Saito, and
  Tahbaz-Salehi}]{carvalho_QJE}
Carvalho VM, Nirei M, Saito YU, Tahbaz-Salehi A (2020) {Supply Chain
  Disruptions: Evidence from the Great East Japan Earthquake*}. The Quarterly
  Journal of Economics 136(2):1255--1321,
  \href{http://dx.doi.org/10.1093/qje/qjaa044}{\doi{10.1093/qje/qjaa044}}

\bibitem[{Dupor(1999)}]{dupor}
Dupor B (1999) Aggregation and irrelevance in multi-sector models. Journal of
  Monetary Economics 43(2):391 -- 409,
  \href{http://dx.doi.org/10.1016/S0304-3932(98)00057-9}{\doi{10.1016/S0304-3932(98)00057-9}}

\bibitem[{Erkel-Rousse and Mirza(2002)}]{cje02}
Erkel-Rousse H, Mirza D (2002) Import price elasticities: Reconsidering the
  evidence. Canadian Journal of Economics 35:282--306,
  \href{http://dx.doi.org/10.1111/1540-5982.00131}{\doi{10.1111/1540-5982.00131}}

\bibitem[{Eslava et~al(2004)Eslava, Haltiwanger, Kugler, and Kugler}]{eslava}
Eslava M, Haltiwanger J, Kugler A, Kugler M (2004) The effects of structural
  reforms on productivity and profitability enhancing reallocation: evidence
  from colombia. Journal of Development Economics 75(2):333--371,
  \href{http://dx.doi.org/https://doi.org/10.1016/j.jdeveco.2004.06.002}{\doi{https://doi.org/10.1016/j.jdeveco.2004.06.002}}

\bibitem[{Foster et~al(2008)Foster, Haltiwanger, and Syverson}]{foster}
Foster L, Haltiwanger J, Syverson C (2008) Reallocation, firm turnover, and
  efficiency: Selection on productivity or profitability? American Economic
  Review 98(1):394--425,
  \href{http://dx.doi.org/10.1257/aer.98.1.394}{\doi{10.1257/aer.98.1.394}}

\bibitem[{{FRED}(2021)}]{fred}
{FRED} (2021) {Federal Reserve Economic Data}.
  \urlprefix\url{https://fred.stlouisfed.org/series/GDP}

\bibitem[{Gabaix(2011)}]{gabaixECTA}
Gabaix X (2011) The granular origins of aggregate fluctuations. Econometrica
  79(3):733--772,
  \href{http://dx.doi.org/10.3982/ECTA8769}{\doi{10.3982/ECTA8769}}

\bibitem[{Hawkins and Simon(1949)}]{hsecta}
Hawkins D, Simon HA (1949) Note: Some conditions of macroeconomic stability.
  Econometrica 17(3/4):245--248,
  \urlprefix\url{http://www.jstor.org/stable/1905526}

\bibitem[{Horvath(1998)}]{horvath98}
Horvath M (1998) Cyclicality and sectoral linkages: Aggregate fluctuations from
  independent sectoral shocks. Review of Economic Dynamics 1(4):781--808,
  \href{http://dx.doi.org/10.1006/redy.1998.0028}{\doi{10.1006/redy.1998.0028}}

\bibitem[{Horvath(2000)}]{horvath}
Horvath M (2000) Sectoral shocks and aggregate fluctuations. Journal of
  Monetary Economics 45(1):69 -- 106,
  \href{http://dx.doi.org/10.1016/S0304-3932(99)00044-6}{\doi{10.1016/S0304-3932(99)00044-6}}

\bibitem[{Hulten(1978)}]{hulten}
Hulten CR (1978) {Growth Accounting with Intermediate Inputs}. The Review of
  Economic Studies 45(3):511--518,
  \href{http://dx.doi.org/10.2307/2297252}{\doi{10.2307/2297252}}

\bibitem[{Hurn et~al(2003)Hurn, Lindsay, and Martin}]{hlm}
Hurn AS, Lindsay KA, Martin VL (2003) On the efficacy of simulated maximum
  likelihood for estimating the parameters of stochastic differential
  equations*. Journal of Time Series Analysis 24(1):45--63,
  \href{http://dx.doi.org/https://doi.org/10.1111/1467-9892.00292}{\doi{https://doi.org/10.1111/1467-9892.00292}}

\bibitem[{{JIP}(2019)}]{jip}
{JIP} (2019) Japan industrial productivity database 2018.
  \urlprefix\url{http://www.rieti.go.jp/en/database/jip.html}

\bibitem[{Kennan(2001)}]{kennan}
Kennan J (2001) Uniqueness of positive fixed points for increasing concave
  functions on {R}n: An elementary result. Review of Economic Dynamics 4(4):893
  -- 899,
  \href{http://dx.doi.org/10.1006/redy.2001.0133}{\doi{10.1006/redy.2001.0133}}

\bibitem[{Krasnosel'ski\u{\i}(1964)}]{kras}
Krasnosel'ski\u{\i} MA (1964) Positive Solutions of Operator Equations.
  Groningen, P. Noordhoff

\bibitem[{Long and Plosser(1983)}]{lpJPE}
Long JB, Plosser CI (1983) Real business cycles. Journal of Political Economy
  91(1):39--69, \href{http://dx.doi.org/10.1086/261128}{\doi{10.1086/261128}}

\bibitem[{Marathe and Ryan(2005)}]{engeco}
Marathe RR, Ryan SM (2005) On the validity of the geometric brownian motion
  assumption. The Engineering Economist 50(2):159--192,
  \href{http://dx.doi.org/10.1080/00137910590949904}{\doi{10.1080/00137910590949904}}

\bibitem[{Miranda-Pinto and Young(2022)}]{MPY}
Miranda-Pinto J, Young ER (2022) Flexibility and frictions in multisector
  models. American Economic Journal: Macroeconomics 14(3):450--80,
  \href{http://dx.doi.org/10.1257/mac.20190097}{\doi{10.1257/mac.20190097}}

\bibitem[{Saito(2004)}]{saito}
Saito M (2004) Armington elasticities in intermediate inputs trade: a problem
  in using multilateral trade data. Canadian Journal of Economics
  37(4):1097--1117,
  \href{http://dx.doi.org/10.1111/j.0008-4085.2004.00262.x}{\doi{10.1111/j.0008-4085.2004.00262.x}}

\bibitem[{Takayama(1985)}]{takayama}
Takayama A (1985) Mathematical Economics. Cambridge University Press

\end{thebibliography}
}
\end{document}